\documentclass[12pt,a4paper]{article}
\usepackage{amssymb} 
\usepackage{amsmath}
\usepackage{mathtools}
\usepackage{amsfonts}    
\usepackage{dsfont}
\usepackage{pdfpages}
\usepackage{verbatim}
\usepackage{tensor}
\usepackage{hyperref}
\usepackage{mathrsfs}
\usepackage[mathscr]{euscript}
\usepackage{tikz}
\usepackage[utf8]{inputenc}
\usepackage[symbol]{footmisc}

\usepackage{tikz-cd}
\usepackage{extarrows}

\newcommand{\ba}{\begin{align}}

\newcommand{\be}{\begin{equation}}
\newcommand{\ee}{\end{equation}}
\def\bd{\begin{tikzpicture}}
\def\ed{\end{tikzpicture}}

\DeclareMathOperator{\SU}{SU}

\DeclareMathOperator*{\OplusInt}{%
\mathchoice%
  {\ooalign{$\displaystyle\oplus$\cr\hidewidth$\displaystyle\int$\hidewidth\cr}}
  {\ooalign{\raisebox{.14\height}{\scalebox{.7}{$\textstyle\oplus$}}\cr\hidewidth$\textstyle\int$\hidewidth\cr}}
  {\ooalign{\raisebox{.2\height}{\scalebox{.6}{$\scriptstyle\oplus$}}\cr$\scriptstyle\int$\cr}}
  {\ooalign{\raisebox{.2\height}{\scalebox{.6}{$\scriptstyle\oplus$}}\cr$\scriptstyle\int$\cr}}
}

\newcommand{\interior}[1]{%
  {\kern0pt#1}^{\mathrm{o}}%
}

\advance\voffset by -1.5cm
\advance\hoffset by -0.8cm
\renewcommand{\thefootnote}{\arabic{footnote}}

\newcommand{\symfootnote}[1]{%
\let\oldthefootnote=\thefootnote%
\stepcounter{mpfootnote}%
\addtocounter{footnote}{-1}%
\renewcommand{\thefootnote}{\fnsymbol{mpfootnote}}%
\footnote{#1}%
\let\thefootnote=\oldthefootnote%
}

\allowdisplaybreaks[1]

\begin{document}
\vspace*{1.5cm}

\begin{center}
{\Large 
{\bf The $\mathfrak{u}(2|2)_1$ WZW model}}
\vspace{2.5cm}

\setcounter{mpfootnote}{2}

{\large Matthias R.\ Gaberdiel}\symfootnote{{\tt E-mail: gaberdiel@itp.phys.ethz.ch}} \\
Institut f{\"u}r Theoretische Physik, ETH Z{\"u}rich\\
CH-8093 Z{\"u}rich, Switzerland\\
\vspace{0.5cm}
{\large and} \\

\setcounter{mpfootnote}{0}

\vspace{0.5cm}
{\large Elia Mazzucchelli}\symfootnote{{\tt E-mail: eliam@mpp.mpg.de}} \\
Max-Plank-Institut f\"ur Physik, Werner-Heisenberg-Institut,\\
D-85748 M\"unchen, Germany \\
\vspace*{0.5cm}

{\bf Abstract}
\end{center}

WZW models based on super Lie algebras play an important role for the description of string theory on AdS spaces. In particular, for the case of ${\rm AdS}_3 \times {\rm S}^3$ with pure NS-NS flux the super Lie algebra of $\mathfrak{psu}(1,1|2)_k$ appears in the hybrid formalism, and higher dimensional AdS spaces can be described in terms of related supergroup cosets. In this paper we study the WZW models based on $\mathfrak{u}(2|2)_1$ and $\mathfrak{psu}(2|2)_1$ that may play a role for the worldsheet theory that is dual to free super Yang-Mills in 4D.

\newpage
\renewcommand{\theequation}{\arabic{section}.\arabic{equation}}

\section{Introduction} \label{sec:intro}

Recently an exactly solvable AdS/CFT duality was found, relating string theory on  ${\rm AdS}_3 \times {\rm S}^3 \times \mathbb{T}^4$ with $k=1$ unit of NS-NS flux to the symmetric orbifold of $\mathbb{T}^4$ \cite{Eberhardt:2018ouy,Eberhardt:2019ywk}, see also \cite{Gaberdiel:2018rqv} for earlier work. The worldsheet theory is best described using the hybrid formalism of \cite{Berkovits:1999im}, for which the 
${\rm AdS}_3 \times {\rm S}^3$ part of the background is captured by a WZW model based on the super Lie algebra $\mathfrak{psu}(1,1|2)_k$. A crucial step in the analysis of \cite{Eberhardt:2018ouy} was to understand the representation theory of $\mathfrak{psu}(1,1|2)_1$ at level $k=1$, for which the only allowed representations are `short' (or `atypical'). This is at the heart of why an NS-NS background, for which one would naively expect a continuous spectrum of long string states, can be dual to a discrete 2d CFT: the shortening condition effectively removes these continuous degrees of freedom. 

It was subsequently suggested that a similar construction could also work for ${\rm AdS}_5 \times {\rm S}^5$ at the point in moduli space where the string theory is dual to free super Yang Mills, and the worldsheet theory that was proposed in \cite{Gaberdiel:2021qbb,Gaberdiel:2021jrv} involves a WZW model based on $\mathfrak{psu}(2,2|4)_k$, again at level $k=1$.\footnote{More generally, string theory on ${\rm AdS}_5 \times {\rm S}^5$ should have  a supercoset description, see e.g.\ \cite{Metsaev:1998it}.} While the description in  \cite{Gaberdiel:2021qbb,Gaberdiel:2021jrv} reproduced the correct spectrum of free ${\cal N}=4$ SYM in 4D, its worldsheet theory exhibits a global $\mathfrak{psu}(2,2|4) \oplus \mathfrak{psu}(2,2|4)$ symmetry. This is a bit surprising, given that the actual ${\cal N}=4$ superconformal symmetry in 4D is described by a single copy of $\mathfrak{psu}(2,2|4)$. Thus somehow the physical state condition of the proposed worldsheet theory of \cite{Gaberdiel:2021qbb,Gaberdiel:2021jrv} must remove half of this symmetry. 

While this may certainly be possible --- and indeed the prescription of \cite{Gaberdiel:2021qbb,Gaberdiel:2021jrv} effectively implements this --- one may wonder whether a more `economical' description of the relevant worldsheet theory could involve a theory with only a smaller symmetry algebra. One natural starting point for such a scenario would be the symmetry\footnote{Strictly speaking, the correct embedding is $\mathfrak{p}\bigl(\mathfrak{su}(2|2) \oplus \mathfrak{su}(2|2) \bigr) \subset \mathfrak{psu}(2,2|4)$.} $\mathfrak{psu}(2|2)  \oplus \mathfrak{psu}(2|2) \subset \mathfrak{psu}(2,2|4)$ that also appears naturally in the BMN set-up and that plays a crucial role for the integrability analysis, see e.g.\ \cite{Beisert:2005tm}. This line of reasoning would then  suggest that the underlying worldsheet theory could involve a WZW model based on $\mathfrak{(ps)u}(2|2)_1$. It is the aim of this paper to analyse this WZW model in detail. In particular, we shall classify and construct all representations of $\mathfrak{u}(2|2)_1$ in detail, and identify a number of modular invariants. While some of the results are quite parallel to what happens for $\mathfrak{psu}(1,1|2)_1$, there are a number of new features that play an important role. First of all, the representations of $\mathfrak{u}(2|2)_1$  carry two additional labels relative to $\mathfrak{psu}(2|2)_1$ (or $\mathfrak{psu}(1,1|2)_1$). More significantly, both $\mathfrak{su}(2)$ subalgebras of  $\mathfrak{u}(2|2)$ correspond to compact groups, and the eigenvalues of the corresponding Cartan generators must therefore be discrete; this changes the nature of the modular invariants quite significantly. We also find, quite intriguingly, modular invariants that are maximally extended and only involve one summand, see e.g.\ eqs.~(\ref{5.14}) and (\ref{L_invariant}). 
\smallskip

Independently of the above string theory considerations, WZW models based on superalgebras have also been studied in their own right. In particular, they provide some of the simplest (and first) examples of logarithmic conformal field theories \cite{Rozansky:1992rx} and exhibit many interesting mathematical challenges related to the presence of reducible but indecomposable modules \cite{Schomerus:2005bf,Gotz:2006qp,Saleur:2006tf}, such as the convergence properties of their characters \cite{Flohr:1995ea,Creutzig:2012sd,Creutzig:2013yca} as well as the complicated structure of their fusion rules \cite{Gaberdiel:2001ny,Ridout:2008nh,Ridout:2010jk}, see \cite{Flohr:2001zs,Gaberdiel:2001tr,Creutzig:2013hma} for some reviews. It is therefore also interesting from this more abstract viewpoint to study another example of this kind.
\bigskip

The paper is organised as follows: In Section~\ref{section_u22} we define the super Lie algebra $\mathfrak{u}(2|2)_k$, as well as its various subalgebras and quotient algebras. We also classify the possible highest weight representations at level $k=1$, see in particular Section~\ref{sec:u22k1}. In Section~\ref{section_free_field_real} we explain a free field realisation for the algebra $\mathfrak{u}(2|2)_1$, and show how the previously identified highest weight representations arise from this free field perspective. Section~\ref{sec:characters} explains how the characters of these representations (as well as their spectrally flowed images) can be calculated, and determines the corresponding modular matrices. In Section~\ref{sec:modinv}  we then identify a number of modular invariants and comment on some of their properties. Section~\ref{sec:concl} contains our conclusions, and there are three appendices which contain some of the more technical material.

\section{The algebra $\mathfrak{u}(2|2)_k$ and its representations}
\setcounter{equation}{0}
\label{section_u22}

Let us begin by studying the super Lie algebra $\mathfrak{u}(2|2)_k$. It contains the bosonic subalgebra 
\begin{equation}\label{u-bosonic}
    \mathfrak{su}(2)_{-k} \oplus \mathfrak{su}(2)_k \oplus \mathfrak{u}(1)_{-k/2} \oplus \mathfrak{u}(1)_{k/2} \ ,
\end{equation}
whose generators we denote by $J^{a}_{n}$, $K^{a}_{n}$ for $a = 3, \pm$ for the two $\mathfrak{su}(2)$ factors, and $U_{n}$ and $V_{n}$ for the two $\mathfrak{u}(1)$ factors, respectively. For the affine $\mathfrak{su}(2)$ generators we use the familiar conventions 
\begin{equation}
\label{su2_k}
\begin{aligned}
    [K^{3}_{m}, K^{3}_{n}] &= \tfrac{k}{2}   m  \delta_{m + m,0}\ ,\\
    [K^{3}_{m},K^{\pm}_{n}] &= \pm K^{\pm}_{m+n} \ , \\
    [K^{+}_{m}, K^{-}_{n}] &= 2  K^{3}_{m+n} + k  m  \delta_{m+n,0}\ ,
\end{aligned}
\end{equation}
while the $J^{a}_{n}$ satisfy the same algebra except that $k$ is replaced by $-k$. Finally, the two $\mathfrak{u}(1)$ generators satisfy 
\begin{equation}\label{UV_commutators}
    [U_{m}, U_{n}]  = - \tfrac{k}{2}  m  \delta_{m+n,0}\ , \qquad 
    [V_{m}, V_{n}] = \tfrac{k}{2}  m  \delta_{m+n,0}\ .
\end{equation}
The eight fermionic generators $S^{\alpha \beta \gamma}_{n}$, where $\alpha,\beta,\gamma \in\{\pm\}$ transform as $2\cdot (\mathbf{2}, \mathbf{2})$ with respect to the two $\mathfrak{su}(2)$ algebras, and they carry charges $\pm \frac{1}{2}$ with respect to the $\mathfrak{u}(1)$ algebras. More specifically, $\alpha$ and $\beta$ are spinor indices with respect to the $J^a_0$ and $K^a_0$ action, respectively, 
\begin{equation}\label{commu}
    \begin{aligned}[t]
    [J^{3}_{m}, S^{\pm \beta \gamma}_{n}] &= \pm \tfrac{1}{2} S^{\pm \beta \gamma}_{m+n} \ , \\
    [J^{\pm}_{m}, S^{ \mp \beta \gamma}_{n}] &= S^{\pm \beta \gamma}_{m+n} \ ,\\
    [J^{\pm}_{m}, S^{ \pm \beta \gamma}_{n}] &= 0 \ ,
    \end{aligned}
    \qquad \qquad
    \begin{aligned}[t]
    [K^{3}_{m}, S^{\alpha \pm \gamma}_{n}] &= \pm \tfrac{1}{2}  S^{\alpha \pm \gamma}_{m+n}\ , \\
    [K^{\pm}_{m}, S^{\alpha \mp \gamma}_{n}] &=  S^{\alpha \pm \gamma}_{m+n}\ ,\\
    [K^{\pm}_{m}, S^{\alpha \pm \gamma}_{n}] &=  0 \ ,
    \end{aligned}
\end{equation}
or more compactly 
\begin{equation}
    \begin{aligned}[t]
    [J^{a}_{m}, S^{\alpha \beta \gamma}_{n}] &= \tfrac{1}{2}   \tensor{(\sigma^{ a})}{^\alpha_{\nu}} \, S^{\nu \beta \gamma}_{m+n}\ ,  \\
    \end{aligned}
    \qquad \qquad
    \begin{aligned}[t]
    [K^{a}_{m}, S^{\alpha \beta \gamma}_{n}] &= \tfrac{1}{2}   \tensor{(\sigma^{ a})}{^\beta_\nu} \, S^{\alpha \nu \gamma}_{m+n}\ , \\
    \end{aligned}
\end{equation}
where the only non-zero components of the Pauli matrices $\sigma^{a}$ are 
\begin{equation}\label{Pauli}
 \tensor{(\sigma^{ +})}{^-_{+}} = 2 \ , \qquad 
\tensor{(\sigma^{ -})}{^+_{-}} = 2  \ , \qquad 
\tensor{(\sigma^{ 3})}{^+_{+}} = - \tensor{(\sigma^{ 3})}{^-_{-}}  =  1  \ , 
\end{equation}
and $a \in \{\pm,3 \}$ is an $\mathfrak{su}(2)$ adjoint index which is raised and lowered by the standard $\mathfrak{su}(2)$-invariant form
\begin{equation}
    \eta_{+-} = \eta_{-+} = \tfrac{1}{2} \ , \qquad \eta_{33} = 2 \ .
\end{equation}
On the other hand, $\gamma$ captures the charges with respect to the two $\mathfrak{u}(1)$ algebras\footnote{$\gamma$ is in fact a spinor index with respect to an outer $\mathfrak{su}(2)$ action.}
\begin{equation}
    [U_{m}, S^{\alpha \beta \gamma}_{n}] = \tfrac{1}{2} \gamma    S^{\alpha \beta \gamma}_{m+n}\ , 
    \qquad \qquad
    [V_{m}, S^{\alpha \beta \gamma}_{n}] = -\tfrac{1}{2} \gamma   S^{\alpha \beta \gamma}_{m+n}\ .    
\end{equation}
It is often convenient to introduce the $\mathfrak{u}(1)$ combinations 
\begin{equation}
    Z_{n} = U_{n} + V_{n} \ , \qquad  Y_{n} = U_{n} - V_{n} \ ,
\end{equation}
so that
\begin{equation}
\label{ZY}
    \begin{aligned}
     &[Z_{m}, J^{a}_{n}] = [Z_{m}, K^{a}_{n}] = 0 = [Y_{m}, J^{a}_{n}] = [Y_{n}, K_{n}^{a}]\ ,\\
    &[Z_{m}, Z_{n}] = [Y_{m}, Y_{n}] = 0\ ,\\
    &[Z_{m}, S_{n}^{\alpha \beta \gamma}] = 0\ ,\\
    &[Y_{m}, Z_{n}] = - k m \delta_{m+n,0}\ , \qquad 
    [Y_{m}, S^{\alpha \beta\gamma}_{n}] = \gamma  S_{n+m}^{\alpha \beta \gamma}\ . 
\end{aligned}
\end{equation}
The anti-commutators between the fermions can be compactly written as
\begin{equation}\label{comm_SS_general}
\begin{split}
    \{S^{\alpha \beta \gamma}_{m}, S^{\mu \nu \rho}_{n}\} = & - \varepsilon ^{\beta  \nu} \varepsilon ^{\gamma  \rho} \tensor{\sigma}{_{ a}^{\alpha \mu}}\, J^{a}_{m+n} + \varepsilon^{\alpha \mu} \varepsilon^{\gamma  \rho} \tensor{\sigma}{_{a}^{\beta \nu}}   K^{a}_{m+n} \\
&    - \, \varepsilon^{\alpha \mu}  \varepsilon^{\beta \nu} \delta^{\gamma, - \rho}  Z_{m+n} +  \varepsilon^{\alpha \mu}  \varepsilon^{\beta \nu} \epsilon^{\gamma \rho} k m  \delta_{m+n,0} \ , 
\end{split}
\end{equation}
where the only non-zero $\sigma_a{}^{\alpha\beta}$ tensors are 
\begin{equation}
( \sigma_{+})^{++} = -1 \ , \qquad 
( \sigma_{-})^{--} = 1 \ , \qquad 
( \sigma_{3})^{+-} = ( \sigma_{3})^{-+} = 1 \ . 
\end{equation}

Note that the $Z_{n}$ modes are central, i.e.\ commute with the $J^a_n$, $K^a_n$, as well as the $S^{\alpha\beta\gamma}_n$, while the modes $Y_{n}$ are those that extend the algebra $\mathfrak{su}(2|2)_{k}$ to $\mathfrak{u}(2|2)_{k}$. This is to say, the subalgebra $\mathfrak{su}(2|2)_{k}$ is described by the above relations provided we restrict to the modes $J^a_n$, $K^a_n$, $Z_n$, as well as the fermionic generators $S^{\alpha\beta\gamma}_n$.  Finally, the algebra $\mathfrak{psu}(2|2)_{k}$ is generated by the modes $J^a_n$, $K^a_n$, $S^{\alpha\beta\gamma}_n$, where we simply drop the $Z_n$ dependent terms in (\ref{comm_SS_general}).

We shall be mainly interested in the above algebras at level $k=1$. Then both $\mathfrak{u}(2|2)_1$ and $\mathfrak{psu}(2|2)_1$ have conformal embeddings, see e.g.\ \cite{Goddard:1987td,Conf_emb}, 
\begin{equation}
\begin{aligned}
        T^{\, \mathfrak{u}(2|2)_{1}} &= T^{\mathfrak{su}(2)_{-1}} + T^{\mathfrak{su}(2)_{1}} + T^{\mathfrak{u}(1)_{-1/2}} + T^{\mathfrak{u}(1)_{1/2}} \ , \\
    T^{\, \mathfrak{psu}(2|2)_{1}} &= T^{\mathfrak{su}(2)_{-1}} + T^{\mathfrak{su}(2)_{1}} \ ,
\end{aligned}
\end{equation}
where $T^{{\mathfrak{g}}}$ denotes the stress-energy tensor of $\mathfrak{g}_k$, which is given by the Sugawara construction. (For $\mathfrak{u}(2|2)_{1}$ we need a generalised Sugawara construction since $\mathfrak{u}(2|2)_{1}$ is not simple, see e.g.\ \cite{Halpern:1989ss,Gaberdiel:2022als}). On the other hand, no such conformal embedding is possible for $\mathfrak{su}(2|2)$ since there is no Sugawara construction for the null-like $\mathfrak{u}(1)$ current $Z_n$. The above conformal embeddings therefore imply that, on (Virasoro) highest weight states, we have 
\begin{equation}\label{conformalembedding}
\begin{aligned}
        L_{0}^{\mathfrak{u}(2|2)_{1}} &= L^{\mathfrak{psu}(2|2)_{1}}_{0} - Z_{0}Y_{0} \ , \\
        L^{\mathfrak{psu}(2|2)_{1}}_{0} &=  C^{\mathfrak{su}(2)_{-1}} + \tfrac{1}{3}C^{\mathfrak{su}(2)_{1}} \ , 
\end{aligned}
\end{equation}
where the $\mathfrak{su}(2)_{\pm 1}$ Casimirs are given by
\begin{equation}
    C^{\mathfrak{su}(2)_{-1}} = J^{3}_{0}  J^{3}_{0} + \tfrac{1}{2}  (J^{+}_{0}  J^{-}_{0} + J^{-}_{0} J^{+}_{0}) \ ,
   \qquad 
    C^{\mathfrak{su}(2)_{1}} = K^{3}_{0}  K^{3}_{0} + \tfrac{1}{2}  (K^{+}_{0}  K^{-}_{0} + K^{-}_{0} K^{+}_{0}) \ .
\end{equation}

The superalgebra $\mathfrak{u}(2|2)$ possesses a \textit{conjugation} automorphism, denoted in the following by $*$, which extends the usual $\mathfrak{su}(2)$ conjugation to the full superalgebra, \be\label{conjugation_aut}
\begin{array}{rclrclrcl}
(J^{3})^* & = & - J^{3}  \ , \qquad & (K^{3})^* & = & - K^{3} \ , \qquad & & & \\
(J^{\pm})^* & = & J^{\mp} \ , \qquad & (K^{\pm})^* & = & K^{\mp} \ ,  \qquad  & (S^{\alpha \beta \gamma})^* & = &   S^{-\alpha , -\beta , -\gamma} \ , \\ 
Z^* & = & -Z \ , \qquad & Y^* & = & -Y \ .  \qquad & & & 
\end{array}
\ee
This automorphism descends to $\mathfrak{su}(2|2)$ and $\mathfrak{psu}(2|2)$, and can also be extended to the corresponding affine superalgebras. 

The affine algebra $\mathfrak{u}(2|2)_k$ also contains automorphisms that are usually referred to as  spectral flow, and that can be defined as follows: for each (Cartan) generator $P_m \in\mathfrak{u}(2|2)_k$ that acts diagonally on all other modes as 
\begin{equation}
{}[P_m, A_n] = \alpha_P(A) \, A_{m+n} + \beta_P(A) \, m \delta_{m+n,0}  \ ,
\end{equation} 
we define the spectral flow automorphism via
\begin{equation}\label{spectralflow}
\sigma_P(A_n) = A_{n+ \alpha_P(A)} +  \beta_P(A) \delta_{n,0} \ .
\end{equation}
The induced action on the Virasoro generators is then
\begin{equation}\label{spectralflow1}
\sigma_P(L_n) = L_n +  P_n + \tfrac{1}{2} \beta_P(P) \delta_{n,0} \ . 
\end{equation}

Each automorphism of the affine superalgebra induces an action on the representations. More specifically, if ${\cal H}$ is any representation of $\mathfrak{u}(2|2)_k$ and $\sigma$ is an automorphism, we define the representation $\sigma\bigl({\cal H}\bigl)$ to consist, as a vector space, of the vectors $[\Phi]^\sigma$, where $\Phi\in {\cal H}$ is arbitrary. (Thus, as vector spaces, these representation spaces are isomorphic.) The action of the $\mathfrak{u}(2|2)_k$ modes on $\sigma\bigl({\cal H}\bigl)$ however differs, since it is twisted by the automorphism, i.e.\ we set 
\begin{equation}
A_n \bigl[\Phi\bigl]^\sigma : = \bigl[ \sigma(A_n) \Phi\bigl]^\sigma \ . 
\end{equation}
We shall call $\mathcal{H}^{*}$ and $\sigma^{w}_P \bigl(\mathcal{H} \bigl)$ the conjugate and $w$'th spectrally flowed modules of $\mathcal{H}$, respectively. (Here $w\in\mathbb{Z}$.)

\subsection{The representations of $\mathfrak{su}(2)$}
\label{su2_representations}

In order to analyse the affine $\mathfrak{u}(2|2)_{k}$ representations, it is necessary to develop first the representation theory of the finite $\mathfrak{u}(2|2)$ Lie superalgebra. Since the bosonic subalgebra is $\mathfrak{su}(2) \oplus \mathfrak{su}(2)\oplus \mathfrak{u}(1) \oplus \mathfrak{u}(1)$, we can decompose representations of $\mathfrak{u}(2|2)$ in terms of multiplets of $\mathfrak{su}(2)$  and $\mathfrak{u}(1)$ representations. 
In the affine algebra, one of the two $\mathfrak{su}(2)$'s appears at level $+k$, while the other has level $-k$. For positive integer $k$, $k \in \mathbb{N}$, the $\mathfrak{su}(2)_{k}$ algebra is integrable, and it possesses only $k+1$ unitary integrable highest weight representations, which are characterised by the 
finite-dimensional spin $\ell \in \tfrac{1}{2}\mathbb{N}$ representation of $\mathfrak{su}(2)$  (with $\ell\leq \frac{k}{2}$)  with respect to which the highest weight states transform. However, the $\mathfrak{su}(2)_{-k}$ model is non-integrable, and thus no such restriction applies for it.  For this reason, we refer to $\mathfrak{su}(2)_{k}$ and $\mathfrak{su}(2)_{-k}$ as the \textit{compact} and \textit{non-compact} factor, respectively, of the bosonic subalgebra of $\mathfrak{u}(2|2)_{k}$. 

Even though the spectrum of the $\mathfrak{su}(2)_{-k}$ theory is continuous, from the perspective of the Lie group, we expect the compactness of $\SU(2)$ to constrain the set of allowed representations to a discrete subset characterised by the property that the magnetic quantum numbers are quantised. For the moment we will ignore this constraint and consider all possible representations of $\mathfrak{su}(2)$, i.e.\ all possible values of the Casimir
\begin{equation}\label{Casimir}
C^{\mathfrak{su}(2)} = j(j+1) \ . 
\end{equation}
In addition, the representation is characterised by the spectrum of the magnetic quantum numbers, i.e.\ the eigenvalues of $J^3$ mod $1$, which we shall parameterise as
\begin{equation}
J^3 = \lambda \ \hbox{mod 1} \qquad \hbox{with $\lambda\in[0,1)$ .}
\end{equation}
Generically, the resulting representation is a {\em continuous} representation, generated by the states of the form $|m\rangle$, $m\in \lambda + \mathbb{Z}$, with\footnote{Obviously there is some choice in how to normalise the states $|m\rangle$, and here we are working with the convention that $J^-$ never vanishes. For generic values of $j$ and $\lambda$ this normalisation is irrelevant, and one could also work with the convention that, say, $J^+$ never vanishes; for special values of $j$, however, these conventions matter and they lead in general to inequivalent representations.}
\begin{equation}
\begin{aligned}\label{su2_continuous_repr}
     J^{3} \, |m\rangle &= m \, |m \rangle \ ,\\ 
     J^{+} \, |m\rangle &=  \bigl( j(j+1) - m(m+1) \bigl)\, |m+1 \rangle \ ,\\
     J^{-} \, |m\rangle &= |m-1 \rangle  \ .\\           
\end{aligned}
\end{equation}
This representation will be denoted by $C^\lambda_j$. Note that $C^{\lambda}_{j} = C^{\lambda}_{-j-1}$, and hence we may assume that $j \geq -1/2$. Moreover, for $j-\lambda \in \mathbb{Z}$, the continuous representation (\ref{su2_continuous_repr}) is not irreducible since 
$J^{+} \, |j\rangle = J^{+} \, |-j-1 \rangle = 0$. The corresponding subrepresentation
\begin{equation}\label{discrete_subrepresentation}
       \{ |j-m \rangle : m \in \mathbb{N}_0\} \cong D^{+}_{j} 
\end{equation}
is then the {\em highest weight discrete representation} $D^+_j$, i.e.\ the representation that is generated from $|j\rangle$ subject to the constraint 
    \begin{equation}
        D^{\pm}_{j} ~:\quad J^{\pm} \, |j\rangle = 0 \quad \text{and} \quad J^{3} \, |j\rangle = j |j\rangle \ .
    \end{equation}
(If instead $J^-$ annihilates $|j\rangle$, the corresponding representation is denoted by $D^-_{j}$,
and  defines a {\em lowest weight discrete representation}, whose Casimir is parametrised as $j(j-1)$.) The complement of (\ref{discrete_subrepresentation}) does not form a subrepresentation of $C^{\lambda}_{j}$; however, the corresponding quotient space does, and we have 
\begin{equation}
        C_{j}^{\lambda} \, \big/ \,  \{ |j-m \rangle : m \in \mathbb{N}_0\} \cong D^{-}_{j+1} \ .
\end{equation}
A more fancy way of saying this is that  there is a non-split short exact sequence
\begin{equation}\label{Csequence}
    0 \longrightarrow D^{+}_{j} \longrightarrow C^{j}_{j} \longrightarrow D_{j+1}^{-} \longrightarrow 0 \qquad \forall \, j \in \mathbb{R} \ .
\end{equation}
Finally, the finite dimensional representations $H_j$ with $j\in \frac{1}{2} \mathbb{N}_0$ are obtained from either $D^\pm_{\pm j}$ by taking the further quotient 
\begin{equation}
     H_j \cong D^{\pm}_{\pm j} \big/ \,  \{ |\mp (j + 1  + m) \rangle : m \in \mathbb{N}_0\} \quad \text{for}~j \in \tfrac{1}{2}\mathbb{N}_0 \ .
\end{equation}
From now on, we will denote by $\mathbf{n}$ for $n \in \mathbb{N}_{>0}$ the $n$-dimensional representation of $\mathfrak{su}(2)$ $H_j$ of spin $j \in \mathbb{N}_0$ with $2j+1 = n$.
Later we will also need the Clebsch-Gordan coefficients of the tensor product of $D^{\pm}_{j}$ and $C^{\lambda}_{j}$ with $\mathbf{2}$. An explicit calculation shows that 
\begin{equation}\label{Clebsch_coeff}
    C^{\lambda}_{j} \otimes \mathbf{2} \cong C^{\lambda+ 1/2}_{j+1/2} \oplus C^{\lambda+1/2}_{j-1/2}\quad \text{and}\quad  D^{\pm}_{j} \otimes \mathbf{2} \cong D^{\pm }_{j+1/2} \oplus D^{\pm}_{j-1/2} \ .
\end{equation}

\subsection{Representations of $\mathfrak{u}(2|2)$}
\label{section_shortening_condition}

Next we want to describe the representations of $\mathfrak{u}(2|2)$ that will appear as the highest weight states of $\mathfrak{u}(2|2)_k$. It is convenient to describe the highest weight states in terms of the representations of the bosonic subalgebra (\ref{u-bosonic}), i.e.\ in terms of the two $\mathfrak{su}(2)$ algebras, together with the relevant $Y_0$ and $Z_0$ eigenvalues, on which the fermionic zero modes $S^{\alpha \beta \gamma}$ of $\mathfrak{u}(2|2)$ generate a $16$-dimensional Clifford module. 
Let us assume that there are highest weight states that transform in the representation $\big( j \, , \mathbf{n} \big)$ with respect to the bosonic subalgebra $\mathfrak{su}(2) \oplus \mathfrak{su}(2)$, where $j \in \mathbb{R}$ labels the spin of the non-compact $\mathfrak{su}(2)$ representation, while ${\bf n}$ is the dimension of the representation of the compact $\mathfrak{su}(2)$. (For the following it actually does not matter whether the representation labelled by $j$ is continuous, discrete, or finite dimensional.) Since the fermionic generators transform as bi-spinors of $\mathfrak{su}(2)\oplus\mathfrak{su}(2)$, a \textit{typical}\footnote{A module over a Lie superalgebra is called \textit{typical} if it is an irreducible Kac module. Otherwise the corresponding Kac module contains non-trivial fermionic singular vectors and the irreducible quotient is called $\textit{atypical}$.} multiplet takes the form 
\begin{equation}\label{huge_multiplet}
  \begin{gathered} 
    \big( j \, , \mathbf{n} \big)    \\
    \big(j + \tfrac{1}{2} \, , \mathbf{n+1} \big) ~~~ \big(j + \tfrac{1}{2} \, , \mathbf{n-1} \big) ~~~ \big(j - \tfrac{1}{2} \, , \mathbf{n+1} \big)~~~\big(j - \tfrac{1}{2} \, , \mathbf{n-1} \big) \\
    \big(j + 1 \, , \mathbf{n} \big)~~~~ ~~~~ \big(j \, , \mathbf{n+2} \big)~~~~~~~~ 2 \, \big(j \, , \mathbf{n} \big) ~~~~~~~~  \big(j  \, , \mathbf{n-2} \big) ~~~~~~~~  \big(j - 1 \, , \mathbf{n} \big) \\
    \big(j + \tfrac{1}{2} \, , \mathbf{n+1} \big) ~~~ \big(j + \tfrac{1}{2} \, , \mathbf{n-1} \big) ~~~ \big(j - \tfrac{1}{2} \, , \mathbf{n+1} \big)~~~\big(j - \tfrac{1}{2} \, , \mathbf{n-1} \big)
     \\ 
    \big( j \, , \mathbf{n} \big)\ . 
  \end{gathered}
\end{equation}
In addition, the multiplet is also characterised by the $Y$- and $Z$-eigenvalue of the states in the top summand $\bigl(j,\mathbf{n}\bigl)_{Y,Z}$. Since $Z$ is central in $\mathfrak{u}(2|2)$, all summands in (\ref{huge_multiplet}) have the same $Z$-eigenvalue. On the other hand, the $Y$-eigenvalues of the different  summands are determined by the eigenvalue of  $\bigl(j,\mathbf{n}\bigl)$ at the top, as well as the commutation relations (\ref{ZY}). We work with the conventions that the highest weight states at the top are annihilated by all fermionic generators of $Y$-charge equal to $+1$, i.e.\ by $S^{\alpha \beta +}_0$. Then going down each row in (\ref{huge_multiplet}) decreases the $Y$-eigenvalue by one. 

We have assumed here that $\mathbf{n}\geq \mathbf{3}$; for $\mathbf{n} = \mathbf{1}$ and $\mathbf{n} = \mathbf{2}$, some shortenings occur, namely for $\mathbf{n} = \mathbf{2}$ we find 
\begin{equation}\label{full_n_2}
  \begin{gathered}
    \big( j \, , \mathbf{2} \big)    \\ 
    \big(j + \tfrac{1}{2} \, , \mathbf{3} \big) ~~~ \big(j + \tfrac{1}{2} \, , \mathbf{1} \big) ~~~ \big(j - \tfrac{1}{2} \, , \mathbf{3} \big)~~~\big(j - \tfrac{1}{2} \, , \mathbf{1} \big) \\
    \big(j + 1 \, , \mathbf{2} \big)~~~~ ~~~~ \big(j \, , 
 \mathbf{4} \big)~~~~~~~~ 2 \, \big(j \, , \mathbf{2} \big) ~~~~~~~~  ~~~~ ~~~~~~~~  \big(j - 1 \, , \mathbf{2} \big) \\
     \big(j + \tfrac{1}{2} \, , \mathbf{3} \big) ~~~ \big(j + \tfrac{1}{2} \, , \mathbf{1} \big) ~~~ \big(j - \tfrac{1}{2} \, , \mathbf{3} \big)~~~\big(j - \tfrac{1}{2} \, , \mathbf{1} \big)
     \\ 
    \big( j \, , \mathbf{2} \big)\ ,
  \end{gathered}
  \end{equation}
while for $\mathbf{n} = \mathbf{1}$ even more representations are missing,
\begin{equation}\label{full_n_1}
  \begin{gathered}
    \big( j \, , \mathbf{1} \big)    \\
    \big(j + \tfrac{1}{2} \, , \mathbf{2} \big) ~~~ \big(j - \tfrac{1}{2} \, , \mathbf{2} \big) \\
    \big(j + 1 \, , \mathbf{1} \big) ~~~~~ \big(j \, , \mathbf{3} \big) ~~~~~  \, \big(j \, , \mathbf{1} \big) ~~~~~ \big(j - 1 \, , \mathbf{1} \big) \\
    \big(j + \tfrac{1}{2} \, , \mathbf{2} \big) ~~~ \big(j - \tfrac{1}{2} \, , \mathbf{2} \big) 
     \\ 
    \big( j \, , \mathbf{1} \big)\ .
  \end{gathered}
\end{equation}

\subsubsection{The highest weight representations at level $k=1$}\label{sec:u22k1}

Below we will be interested in the affine algebra $\mathfrak{u}(2|2)_{k}$ at level $k = 1$. Then the second bosonic $\mathfrak{su}(2)_{k}$ factor also has level $k = 1$, and as a consequence, the highest weight states are only allowed to transform in the $\mathbf{1}$ and $\mathbf{2}$ representations of $\mathfrak{su}(2)$.\footnote{In this section we are discussing the representations of the finite-dimensional Lie superalgebra $\mathfrak{u}(2|2)$ in which the highest weight states of the corresponding affine representation transform.} Hence, it is clear that all the long representations we have presented above are not allowed at $k = 1$. Let us therefore look systematically for short multiplets. This amounts to determining the conditions under which, e.g.\ the $\big(j \, , \mathbf{3} \big)$ summand in (\ref{full_n_1}) can be quotiented out. This turns out to be possible provided that the $Z_0$-eigenvalue takes a specific value. More specifically, the allowed multiplets will be of the form -- this just corresponds to one of the outer edges of the full diamond in (\ref{full_n_1}) 
\begin{equation}\label{shortrep}
\bigl(j,\mathbf{1} \bigr)_{Y+1,\,Z} \oplus \bigl(j-\tfrac{1}{2},\mathbf{2} \bigr)_{Y,\,Z} \oplus \bigl(j-1,\mathbf{1} \bigr)_{Y-1,\,Z} \ , \qquad \hbox{with $Z=j$ \ ,}
\end{equation}
or 
\begin{equation}\label{shortrep1}
\bigl(j,\mathbf{1} \bigr)_{Y-1,\,Z} \oplus \bigl(j-\tfrac{1}{2},\mathbf{2} \bigr)_{Y,\,Z} \oplus \bigl(j-1,\mathbf{1} \bigr)_{Y+1,\,Z} \ , \qquad \ \hbox{with $Z=-j$ \ .}
\end{equation}
Here the indices $(Y,Z)$ denote the $Y_0$- and $Z_0$-eigenvalues of the different summands. One way to see this is to use the conformal embedding of eq.~(\ref{conformalembedding}): all summands need to have the same $ L_{0}^{\mathfrak{u}(2|2)_{1}}$-eigenvalue, and this implies that 
\be
j(j+1) - Z(Y \pm 1) = (j-\tfrac{1}{2})(j+\tfrac{1}{2}) + \tfrac{1}{4} - Z Y = (j-1) j - Z (Y\mp 1) \ , 
\ee
from which it follows that $Z=\pm j$. Using that $j$ and $-(j+1)$ give rise to the same Casimir $C^{\mathfrak{su}(2)}$, it is easy to see, upon identifying $\bigl(Z,\mathbf{1} \bigl)_{Y+1,\,Z} \cong \bigl(-Z-1,\mathbf{1} \bigl)_{Y+1,\,Z}$ that the two solutions (\ref{shortrep}) and (\ref{shortrep1}) are in fact the same. Note that in the case of continuous $\mathfrak{su}(2)$ representations there is also the parameter $\lambda$,  that does not enter into this shortening analysis, except that the different terms need to respect (\ref{Clebsch_coeff}). 

Since the analysis only depends on the value of the Casimir, it applies to all the different (highest weight) representations of Section~\ref{su2_representations} with only minor modifications: for $j$ labelling finite-dimensional $\mathfrak{su}(2)$ representations, only the terms involving non-negative spin appear in the multiplets (\ref{shortrep}) and (\ref{shortrep1}), while for $j \in \tfrac{1}{2}\mathbb{Z}$ labelling highest weight discrete $\mathfrak{su}(2)$ representations, only the solution $j=-|Z|$ is allowed, with an extra shortening for $Z=0$. We summarise these short multiplets, together with an alternative derivation of the shortening condition, in Appendix~\ref{appendix_shortening}.

For the representations of $\mathfrak{psu}(2|2)_1$ we need to impose $Z=0$, and then the solution takes the form 
\be
\bigl(0,\mathbf{1} \bigr)_{Y-1,0} \oplus \bigl(-\tfrac{1}{2},\mathbf{2} \bigr)_{Y,0} \oplus \bigl(-1,\mathbf{1} \bigr)_{Y+1,0} \ , 
\ee
where $j=0$ and $j=-1$ describe both a representation with vanishing Casimir, while the middle representation has $C^{\mathfrak{su}(2)} = - \frac{1}{4}$. This agrees with what was found for $\mathfrak{psu}(1,1|2)_1$ in \cite{Eberhardt:2018ouy}, see also \cite{Gaberdiel:2011vf}.

\section{Free field realisation of $\mathfrak{u}(2|2)_1$ }\label{section_free_field_real}
\setcounter{equation}{0}

The superalgebra $\mathfrak{u}(2|2)_1$ actually has a free field realisation in terms of symplectic bosons and fermions, and thus we can also understand the above results directly from that perspective. We denote the symplectic bosons by $(\lambda^{\alpha}, \mu^{\dagger}_{\alpha})$ and the complex fermions by $(\psi^{\alpha}, \psi^{\dagger}_{\alpha})$ where $\alpha = 1,2$, and they satisfy the (anti-)commutation relations
\begin{equation}
    [\lambda^{\alpha}_{r}, (\mu^{\dagger}_{\beta})_{s}] = \delta^{\alpha}_{\beta} \, \delta_{r,-s}\ , ~~~~~~\{\psi^{\alpha}_{r}, (\psi^{\dagger}_{\beta})_{s}\} = \delta^{\alpha}_{\beta} \, \delta_{r,-s} \ .
\end{equation}
We combine these fields as $Y_J = (\mu^{\dagger}_{\alpha}, \psi^{\dagger}_{\beta})$ and $X^{I} = (\lambda^{\alpha}, \psi^{\beta})$, and then consider the normal ordered bilinears
\begin{equation}
    \tensor{J}{^I_J} = Y_{J} \, X^{I} \ .
\end{equation}
These fields generate the superalgebra $\mathfrak{u}(2|2)_{1}$. The central generator is 
\begin{equation} 
Z = \tfrac{1}{2} \, Y_{I} Z^{I} \ ,
\end{equation}
and it will play an important role in the following. The other generators are explicitly
\begin{equation}\label{u22_free_fields}
    \begin{aligned}[t]
    J^{3} &= \tfrac{1}{2}\bigl(\mu^{\dagger}_{2} \, \lambda^{2} - \mu^{\dagger}_{1} \, \lambda^{1}\bigl)  \  , \\
    J^{+} &= \mu^{\dagger}_{2} \, \lambda^{1} \  , \quad J^{-} = \mu^{\dagger}_{1} \, \lambda^{2} \ , \\
    U &= \tfrac{1}{2}\bigl(\mu^{\dagger}_{2} \, \lambda^{2} + \mu^{\dagger}_{1} \, \lambda^{1}\bigl) \  , \\ 
    S^{\alpha \beta +} &=  \beta \, \mu^{\dagger}_{\alpha} \, \psi^{\beta} \ , 
    \end{aligned}
    \qquad 
    \begin{aligned}[t]
    K^{3} &= \tfrac{1}{2}\bigl(\psi^{\dagger}_{2} \, \psi^{2} - \psi^{\dagger}_{1} \, \psi^{1}\bigl) \  , \\
    K^{+} &= \psi^{\dagger}_{2} \, \psi^{1}  \  , \quad  K^{-} = \psi^{\dagger}_{1} \, \psi^{2} \ ,  \\
    V &= \tfrac{1}{2} \bigl(\psi^{\dagger}_{2} \, \psi^{2} + \psi^{\dagger}_{1} \, \psi^{1} \bigl) \ ,\\
    S^{\alpha \beta -} &= \alpha  \, \lambda^{\alpha} \, \psi^{\dagger}_{\beta} \ ,
    \end{aligned}
\end{equation}
where $\alpha,\beta = \pm$, and we identify the indices as $+ \equiv 2$, $- \equiv 1$ for $(\mu^{\dagger}_{\alpha},\psi^{\dagger}_{\alpha})$ and as $+ \equiv 1$, $- \equiv 2$ for $(\lambda^{\alpha},\psi^{\alpha})$. Also, normal ordering is to be understood where relevant. We mention that the $Z$-neutral bilinears of the four symplectic fermions give rise to $\mathfrak{u}(2)_{-1}$, while the $Z$-neutral bilinears of the two complex (four real) fermions form $\mathfrak{u}(2)_{1}$.

\subsection{Spectral flow}\label{sec:sf}

As we have seen in eq.~(\ref{spectralflow}) the $\mathfrak{u}(2|2)_1$ algebra possesses various spectral flow automorphisms, and it is also possible to describe them directly in terms of the free fields. To this end, we define automorphisms for the symplectic bosons as
\begin{equation}\label{spec_flow_bos}
    \sigma^{(\alpha)}_{b} \bigl(\lambda_{r}^{\alpha} \bigl) = \lambda^{\alpha}_{r + \frac{1}{2}} \ ,~~~~ ~~~~ \sigma^{(\alpha)}_{b}\bigl((\mu^{\dagger}_{\alpha})_{r}\bigl) = (\mu^{\dagger}_{\alpha})_{r - \frac{1}{2}} \ ,
\end{equation}
and for the fermions as
\begin{equation}
    \sigma^{(\alpha)}_{f}\bigl(\psi_{r}^{\alpha}\bigl) = \psi^{\alpha}_{r + \frac{1}{2}} \ ,~~~~ ~~~~
    \sigma^{(\alpha)}_{f}\bigl((\psi^{\dagger}_{\alpha})_{r}\bigl) = (\psi^{\dagger}_{\alpha})_{r - \frac{1}{2}}\ ,
\end{equation}
where in each case $\alpha = 1,2$. The $\mathfrak{u}(2|2)_1$ spectral flows can then be described in terms of these free field automorphisms. For example, the spectral flows (\ref{spectralflow}) associated to $P=J^3$ and $P=U$ are obtained as 
\begin{equation}\label{flows_bos}
\sigma_{J^3} = \sigma^{(1)}_{b} \circ \bigl(\sigma^{(2)}\bigl)_{b}^{-1} \ , \qquad 
\sigma_{U} = \bigl( \sigma^{(1)}_{b}\bigr)^{-1} \circ \bigl(\sigma^{(2)}\bigl)_{b}^{-1} \ ,
\end{equation}
while for $P=K^3$ and $P=V$ we find 
\begin{equation}\label{flows_ferm}
\sigma_{K^3} = \sigma^{(1)}_{f} \circ \bigl(\sigma^{(2)}\bigl)_{f}^{-1} \ , \qquad 
\sigma_{V} = \sigma^{(1)}_{f}\circ \sigma^{(2)}_{f} \ ,
\end{equation}
and 
\begin{equation}\label{sigma_Z}
    \sigma_{Z} = \sigma_{U} \circ \sigma_{V} \ 
\end{equation}
is the spectral flow associated to the Cartan generator $Z$, which according to (\ref{spectralflow}) simply shifts $Y_0$ and $L_0$ by 
\begin{equation}\label{sigma_Z_flow}
\sigma^{w}_Z\bigl(Y_0\bigl) = Y_0 - w \ , \qquad \sigma^{w}_Z\bigl(L_0\bigl)  = L_0 + w Z_0 \ . 
\end{equation}
Note that (\ref{sigma_Z_flow}) defines an automorphism of $\mathfrak{u}(2|2)_1$ for every $w \in \mathbb{R}$. For the following we will also need the spectral flow defined by 
\begin{equation}\label{sigma}
\sigma = \sigma_{J^3} \circ \sigma_{K^3} \ , 
\end{equation}
and it acts on the generators as
\begin{equation}\label{spec_flow_sigma}
    \begin{aligned}[t]
        \sigma^{w}\bigl(J^{3}_{m}\bigl) &= J^{3}_{m} - \tfrac{w}{2} \, \delta_{m,0} \ , \\
    \sigma^{w}\bigl(J^{\pm}_{m}\bigl) &= J^{\pm}_{m \pm w} \ ,\\ 
    \sigma^{w}\bigl(S^{\alpha \beta \gamma}_{m}\bigl) &= S^{\alpha \beta \gamma}_{m + \frac{w}{2}   (\alpha + \beta)} \ , \\
    \end{aligned}
    \qquad \qquad
    \begin{aligned}[t]
    \sigma^{w}\bigl(K^{3}_{m}\bigl) &= K^{3}_{m} + \tfrac{w}{2} \, \delta_{m,0} \ , \\
    \sigma^{w}\bigl(K^{\pm}_{m}\bigl) &= K^{\pm}_{m \pm w} \ , \\
    \sigma^{w}\bigl(L_{0}\bigl) &= L_{0} + w \, (J^{3}_{0} + K^{3}_{0}) \  , 
    \end{aligned}
\end{equation}
while leaving $Y_0$ and $Z_0$ invariant. Finally, we define $\rho$ to be the automorphism  
\begin{equation}\label{rho_automorphism}
    \rho = \sigma_{K^3} \circ \sigma_V \ ,
\end{equation} 
which acts on the generators as
\begin{equation}\label{rho_action}
    \begin{aligned}[t]
    \rho^{w}\bigl(K^{3}_{m}\bigl) &= K^{3}_{m} + \tfrac{w}{2} \, \delta_{m,0} \ , \\
    \rho^{w}\bigl(K^{\pm}_{m}\bigl) &= K^{\pm}_{m \pm w} \ , \\
   \rho^{w}\bigl(S^{\alpha \beta \gamma}_{m}\bigl) &= S^{\alpha \beta \gamma}_{m + \frac{w}{2}    (\gamma + \beta)} \ , \\
    \end{aligned}
    \qquad \qquad
    \begin{aligned}[t]
    \rho^{w}\bigl(Z_{m}\bigl) &= Z_{m} + \tfrac{w}{2} \, \delta_{m,0} \  , \\
    \rho^{w}\bigl(Y_{m}\bigl) &= Y_{m} - \tfrac{w}{2} \, \delta_{m,0} \  , \\
    \rho^{w}\bigl(L_{0}\bigl) &= L_{0} + w \, K^{3}_{0} + \tfrac{w}{2} \, ( Z_{0} - Y_{0}) + \tfrac{w^2}{2} \ .
    \end{aligned}
\end{equation}

\subsection{The representations in the free field realisation}\label{sect_free_field_repr}

Next we want to discuss how the above representations, see eq.~(\ref{shortrep}),  arise from the free field perspective. First of all, if all fields are half-integer moded (NS sector), then the corresponding highest weight representation is simply the vacuum representation, for which there is a unique highest weight state that transforms trivially, i.e.\ it is annihilated by all zero modes.\footnote{This represents a degenerate case of the analysis above where $j=0$, and the additional terms in eq.~(\ref{shortrep}) are absent.} The vacuum representation is an atypical representation, and it therefore does not actually appear by itself in the WZW spectrum \cite{Quella:2007hr}. (Instead, it will be part of its projective cover, see Appendix \ref{app_proj_cov}.)

In order to describe the interesting representations, we therefore need to consider the R sector, in which all the free fields are integer moded. The R sector contains zero modes of the symplectic bosons, and as such, the space of highest states will be infinite dimensional. In order to describe these highest weight states, let us introduce the occupation numbers $|m_1, m_2 \rangle$, and define the action of the symplectic boson zero modes as, see also \cite{Dei:2020zui}
\begin{equation}\label{sympl_bos_zero}
     \begin{aligned}[t]
     \lambda^{1}_{0} \, |m_{1}, m_{2} \rangle & = 2  m_{1} \, | m_{1} - \tfrac{1}{2}, m_{2} \rangle \ , \\
    \lambda^{2}_{0} \, |m_{1}, m_{2} \rangle & = 2  m_{2} \,   | m_{1} , m_{2} - \tfrac{1}{2} \rangle \ ,
    \end{aligned}
    \qquad \qquad
    \begin{aligned}[t]
    (\mu^{\dagger}_{1})_{0} \, |m_{1}, m_{2} \rangle & = | m_{1} + \tfrac{1}{2}, m_{2} \rangle \ , \\
    (\mu^{\dagger}_{2})_{0} \, |m_{1}, m_{2} \rangle & = | m_{1} , m_{2} + \tfrac{1}{2} \rangle \ ,
    \end{aligned}
\end{equation}
where $m_{i} \in \tfrac{1}{2}\mathbb{Z} + \delta_{i}$ with $\delta_{1}, \delta_{2} \in \mathbb{R}/\tfrac{1}{2}\mathbb{Z} \cong [0,\tfrac{1}{2})$. As in the case of eq.~(\ref{su2_continuous_repr}), we have made a choice here by taking $(\mu^{\dagger}_{i})_{0}$ to never vanish; for generic values of $\delta_{i}$, this choice is irrelevant, but for $m_1,m_2 \in \tfrac{1}{2}\mathbb{N}_0$, it does matter. Thus, for $m_1,m_2 \in \tfrac{1}{2}\mathbb{N}_0$ there are really four different R sectors that are characterised by which generator of each pair $(\lambda^\alpha,\mu^\dagger_\alpha)$ for $\alpha=1,2$ always acts non-trivially.
With the above convention, the $\mathfrak{u}(2)_{-1}$ generators then give\footnote{We use the usual normal ordering convention that positive modes stand to the right of negative modes. Furthermore, we define $: a_{0}b_{0} : \, = \tfrac{1}{2} ( a_{0}b_{0} \pm b_{0} a_{0} )$, where the sign depends on whether $a$ and $b$ are both fermions or not.}
\begin{equation}\label{J_U_action}
    \begin{aligned}
     J^{3}_{0} \, | m_{1}, m_{2} \rangle &= (m_{2} - m_{1} ) \,  | m_{1}, m_{2} \rangle \ , \\
     J^{+}_{0} \, | m_{1}, m_{2} \rangle &= 2  m_{1} \,   | m_{1} - \tfrac{1}{2}, m_{2} + \tfrac{1}{2} \rangle \ , \\
     J^{-}_{0} \, | m_{1}, m_{2} \rangle &= 2   m_{2} \, | m_{1} + \tfrac{1}{2}, m_{2} - \tfrac{1}{2} \rangle \ , \\
     U_{0} \, | m_{1}, m_{2} \rangle &= (m_{1} + m_{2} + \tfrac{1}{2} ) \, | m_{1}, m_{2} \rangle \ , \\
    \end{aligned}
\end{equation}
and the $\mathfrak{su}(2)_{-1}$ Casimir equals 
\begin{equation}
    C^{\mathfrak{su}(2)} = J^{3}_{0}  J^{3}_{0} + \tfrac{1}{2}  (J^{+}_{0}  J^{-}_{0} + J^{-}_{0} J^{+}_{0}) = j  (j + 1) \qquad \hbox{with $j = m_{1} + m_{2}$}\ .
\end{equation}
For the action of the fermionic zero modes we define
\begin{equation}
\label{ferm_zero_modes}
    \psi^{a}_{0} \, |m_{1}, m_{2} \rangle = 0  ~~~~\text{for}~a = 1,2 \ ,
\end{equation}
such that the action of the creation operators $(\psi^{\dagger}_{a})_0$ with $a=1,2$ leads to a $4$-dimensional Clifford module; with respect to $\mathfrak{su}(2)_{1}$, this decomposes into two singlet states
\begin{equation}
    2 \, \cdot \, \mathbf{1} : ~~~~~~ |m_{1}, m_{2}\rangle ~~~~\text{and} ~~~~ (\psi^{\dagger}_{2})_{0} \, (\psi^{\dagger}_{1})_{0} \, |m_{1}, m_{2}\rangle \ ,
\end{equation}
as well as a doublet spanned by
\begin{equation}
    \mathbf{2} : ~~~~~~ (\psi^{\dagger}_{2})_{0} \, |m_{1}, m_{2}\rangle ~~~~\text{and} ~~~~ (\psi^{\dagger}_{1})_{0} \, |m_{1}, m_{2}\rangle \ .
\end{equation}
For $a \neq b$ we can compute,
\begin{equation}
    \begin{aligned}[t]
    (\psi^{\dagger}_{a} \, \psi^{b})_{0} \, |m_{1}, m_{2} \rangle = 0  ~~~~\text{and}~~~~\\
    \end{aligned}
    \begin{aligned}[t]
    (\psi^{\dagger}_{a} \, \psi^{a})_{0} \, |m_{1}, m_{2} \rangle = -\tfrac{1}{2} \, |m_{1}, m_{2} \rangle \ ,
    \end{aligned}
\end{equation}
and hence 
\begin{equation}
    V_{0} \, | m_{1}, m_2 \rangle = - \tfrac{1}{2} \, | m_{1}, m_2 \rangle \  . 
\end{equation}
With our convention for the action of $U_0$ this then leads to 
\begin{equation}
    Z_{0} \, |m_{1}, m_2 \rangle = j \, |m_{1}, m_2 \rangle  ~~~~\text{and}~~~~Y_{0} \, |m_{1}, m_2 \rangle = (j + 1) \, |m_{1}, m_2 \rangle \ ,
\end{equation}
and thus $Z_0, Y_0 \in \tfrac{1}{2}\mathbb{Z} + \delta_{1} + \delta_{2}\,$. For $Z \in \mathbb{R}$ and $\lambda \in [0,1)$ we denote by $\mathcal{C}^{\lambda}_{Z,Z}$ the R sector for fixed $Z_{0} = Z $ defined by (\ref{sympl_bos_zero}) and (\ref{ferm_zero_modes}) with 
\begin{equation}\label{lambda}
\begin{aligned}
        \delta_{1} = \frac{Z-\lambda}{2} ~\text{mod} \, \tfrac{1}{2} ~~~~\text{and}~~~~
           \delta_{2} = \frac{Z+\lambda}{2} ~~\text{mod} \, \tfrac{1}{2} \ .
\end{aligned}
\end{equation}
It is straightforward to check that $\mathcal{C}^{\lambda}_{Z,Z}$ is an affine $\mathfrak{u}(2|2)_{1}$ highest weight representation whose highest weight states transform in the $\mathfrak{u}(2|2)$ multiplet
\begin{equation}\label{short_multiplet} 
    \big( C_{Z}^{\lambda} \, , \mathbf{1} \big)_{Y+1, \, Z} \oplus
    \big(C_{Z - \frac{1}{2}}^{ \lambda - \frac{1}{2}} \, , \mathbf{2} \big)_{Y, \, Z} \oplus \big(C_{Z-1 }^{ \lambda} \, , \mathbf{1} \big)_{Y-1, \, Z} \ ,
\end{equation}
with $Y=Z$, where for each summand we specified the $Y_{0}$- and $Z_{0}$-eigenvalues respectively. More generally, we denote the affine $\mathfrak{u}(2|2)_1$ representation whose highest weight states transform as (\ref{short_multiplet}) by $\mathcal{C}^{\lambda}_{Y,Z}$. It is easy to check that $\bigl(\mathcal{C}^{\lambda}_{Y,Z} \bigl)^{*} \cong \mathcal{C}^{-\lambda}_{-Y,-Z}$.

The eigenvalues of $U_0$ and $V_0$ in (\ref{J_U_action}) depend on our specific normal ordering convention, and this affects the eigenvalue of~$Y_0 = U_0 - V_0$.\footnote{The eigenvalue of $Z_0 = U_0 + V_0$, on the other hand, is unambiguous since the $Z_n$ generators appear on the right-hand-side of the anti-commutator (\ref{comm_SS_general}).} Thus, depending on the choice of these normal ordering conventions, we can also obtain other eigenvalues of~$Y_0$. In this sense, the free field realisation accounts for all representations of the form (\ref{short_multiplet}), not just those for which $Y=Z$. In particular, this construction therefore gives rise to the affine representations associated to the short multiplets found in (\ref{shortrep}) and (\ref{shortrep1}). 

We should also mention that for the case of $\mathfrak{psu}(2|2)_1$, we need to remove the generators associated to $Y_n$ and $Z_n$ and hence set $Z_0=0$. The resulting representation  will be denoted by  $\mathscr{F}_{\lambda}$, where 
\begin{equation}\label{F_lambda}
  \mathscr{F}_{\lambda} ~~~~:~~~~ 
    \big( C_{0}^{\lambda} \ , \mathbf{1} \big) \oplus
    \big(C_{ - \frac{1}{2}}^{ \lambda - \frac{1}{2}} \, , \mathbf{2} \big)\oplus \big(C_{-1 }^{ \lambda} \ , \mathbf{1} \big) \ ,
\end{equation}
and relative to the  conventions of \cite{Eberhardt:2018ouy} we have shifted here $\lambda$ by $\tfrac{1}{2}$.

\subsection{Composition series of indecomposables}

If $\lambda=Z$ mod $1$, i.e.\ for $\delta_1=0$, the $\mathfrak{u}(2|2)$ representation (\ref{short_multiplet}) is reducible but indecomposable, reflecting the indecomposability of the underlying non-compact $\mathfrak{su}(2)$ representation, see eq.~(\ref{Csequence}). In the affine case, the structure is a little bit more complicated: the analogue of eq.~(\ref{Csequence}) only holds for $Z\notin \frac{1}{2}\mathbb{Z}$,
\be
\mathcal{D}^{+}_{Y,Z} \subset \mathcal{C}_{Y,Z}^{Z} \quad \hbox{with} \quad 
\mathcal{C}_{Y,Z}^{Z} \bigl/  \mathcal{D}^{+}_{Y,Z} \cong \mathcal{D}^{-}_{Y,Z} \qquad \text{for}~ Z \notin \tfrac{1}{2}\mathbb{Z} \ , 
\ee
while for $Z \in \tfrac{1}{2}\mathbb{Z}$ we have 
\be
\mathcal{M}_1 \subset \mathcal{M}_2 \subset  \mathcal{M}_3 \subset \mathcal{C}_{Y,Z}^{Z}  
\ee
with
\begin{equation}\label{R_strucutre_diagr}
\begin{aligned}
& \quad \mathcal{M}_1  \cong   \mathcal{D}^{+}_{Y,Z} 
\ , \quad \qquad \quad  \mathcal{M}_2 \bigl/  \mathcal{M}_1  \cong \mathcal{H}_{Y + \, \text{sgn}(Z),Z} 
 \ , \\
&\quad \mathcal{M}_3 \bigl/ \mathcal{M}_2   \cong   \mathcal{D}^{-}_{Y,Z} 
\ , \qquad 
\mathcal{C}_{Y,Z}^{Z}  \bigl/ \mathcal{M}_3  \cong  \mathcal{H}_{Y- \, \text{sgn}(Z),Z}   \ , \\
\end{aligned}
\end{equation}
where $\mathcal{D}^{\pm}_{Y,Z}$ and ${\cal H}_{Y,Z}$ are the $\mathfrak{u}(2|2)_1$ affine highest weight representations built upon the discrete and finite-dimensional $\mathfrak{su}(2)$ representations respectively, see eqs.~(\ref{short_n_1_highest_discrete}) and (\ref{short_n_1}) for more details, and we define $\text{sgn}(0) =1$. Furthermore, the spectral flow $\sigma$ is defined in eq.~(\ref{spec_flow_sigma}).

One way to derive the composition series (\ref{R_strucutre_diagr}) is to note that the statement for $Y=Z=0$ (with $\lambda=0$) follows directly from the analysis of $\mathscr{F}_{0}$ in \cite[Appendix B]{Eberhardt:2018ouy}. The general case can then be deduced by functoriality, namely by applying iteratively the automorphisms $\rho $ of (\ref{rho_automorphism}) and $\sigma_Z$ of (\ref{sigma_Z}) to the composition series of $\mathcal{C}_{0,0}^{0}$.
(We are relying here on the fact that composition series are preserved under the action of automorphisms, as the latter preserve subspace inclusions and short exact sequences.) More specifically,  each component of the composition series of $\mathcal{C}^{0}_{0,0}$ is mapped to the corresponding component of the composition series of $\mathcal{C}^{Z}_{Z,Z}$ under $\rho^{ 2Z}$. Finally, the result for general~$Y$ is obtained by considering the automorphism $\sigma_Z$ (which commutes with $\rho$).

It is maybe worth pointing out that while, generically, spectral flows map highest weight representations to representations that are not highest weight, there are a few important exceptions. In particular, we have 
\begin{equation}\label{spec_flows_iso1}
    \sigma^{\pm 1}\bigl(\mathcal{H}_{Y,Z}\bigl) \cong \mathcal{D}^{\pm}_{Y,Z}  \ ,
\end{equation}
as well as 
\begin{equation}\label{spec_flow_iso_rho}
\begin{aligned}
\rho \bigl(\mathcal{H}_{Y,Z} \bigl) &\cong \mathcal{H}_{Y + \frac{1}{2}, Z + \frac{1}{2}} \ , \qquad \rho \bigl(\mathcal{D}^{\pm}_{Y,Z} \bigl) \cong \mathcal{D}^{\pm}_{Y + \frac{1}{2}, Z + \frac{1}{2}} \ , \qquad \rho \bigl(\mathcal{C}^{Z}_{Y,Z} \bigl) \cong \mathcal{C}^{Z+\frac{1}{2}}_{Y + \frac{1}{2},Z+\frac{1}{2}}  \ ,   
\end{aligned}
\end{equation}
for every $Y \in \mathbb{R}$ and $Z \in \frac{1}{2}\mathbb{Z}$. In order to explain how to derive these identities, let us spell this out for the case $\sigma \bigl( \mathcal{H}_{Y,Z} \bigl ) \cong \mathcal{D}^{+}_{Y,Z}$  --- the arguments for the other cases are similar. 
Suppose first that $Z=0$, and denote by~$|0\rangle $ the highest weight state in $(H_0 \,, \mathbf{1})_{Y ,\,0}$ in $\mathcal{H}_{Y,0}$, see (\ref{short_n_1}), which is annihilated by all the fermionic zero modes $S^{\alpha \beta \gamma}_0$. One easily verifies that the image of $|0\rangle$ under~$\sigma$ transforms in $(D^{+}_{-1/2}, \mathbf{2})_{Y,\,0}$ with respect to the zero modes of the bosonic subalgebra, and  that it is annihilated by all the fermionic zero modes $S^{\alpha\beta\gamma}_0$ except for  $\alpha=\beta=-$, thus generating $\mathcal{D}^{+}_{Y,0}$, see (\ref{short_n_1_highest_discrete}). For $Z > 0$ --- the case $Z<0$ can be analysed similarly --- we consider the state $ |-Z + \tfrac{1}{2} ,\downarrow \rangle $, which is the bosonic lowest weight state in $(H_{Z-1/2} \,, \mathbf{2})_{Y-1,\,Z}$ that is annihilated by the fermionic zero modes $S^{\alpha\beta\gamma}_0$ with $\beta=-$. Then, its spectrally flowed image under $\sigma$ transforms as $(D^{+}_{-Z} \,, \mathbf{1})_{Y-1,\,Z}$ with respect to the bosonic zero modes. It is also annihilated by all positive fermion modes as well as the zero modes $S^{\alpha\beta\gamma}_0$ with $\alpha=-$. This state therefore generates the affine representation $\mathcal{D}^{+}_{Y,Z}$, see (\ref{short_n_1_highest_discrete}).

\section{Characters and modular properties}\label{sec:characters}
\setcounter{equation}{0}

In order to find the modular invariants of $\mathfrak{u}(2|2)_1$, the next step is to determine the characters of the above R sector representations,  and to calculate their modular transformation properties. The calculation proceeds similarly to what was done in Appendix~C of \cite{Eberhardt:2018ouy}, except that now we do not impose that the central term $Z_0=Z$ vanishes. The contribution of the fermions is as in \cite[(C.31)]{Eberhardt:2018ouy}
\be
 \frac{\vartheta_2(z;2\tau) \vartheta_3(\mu-\nu;2\tau) + \vartheta_3(z;2\tau) \vartheta_2(\mu-\nu;2\tau)}{\eta(\tau)^2} \ , 
\ee
where, as usual, $q=e^{2\pi i\tau}$, while $\mu$ and $\nu$ are the chemical potentials associated to $Z_0$ and $Y_0$, and $z$ is the one associated to $K^{3}_0$. (The fermions only depend on $Z_0 - Y_0 = 2 V_0$, i.e.\ the character only depends on $\mu-\nu$.) 
For the bosons we have similarly, see \cite[(C.32)]{Eberhardt:2018ouy} 
\be
\begin{aligned}\label{char_sympl_bos}
& q^{-\frac{1}{6}} \sum_{m_1 \in \delta_1 + \frac{1}{2} \mathbb{Z}} \ \sum_{m_2 \in \delta_2 + \frac{1}{2} \mathbb{Z}} \, x^{m_2-m_1 } \, e^{2\pi i (\mu+\nu) (m_1 + m_2 +\frac{1}{2})} \, \prod_{n=1}^{\infty} \prod_{a,b = \pm \frac{1}{2}} \frac{1}{1-e^{2\pi i a (\mu+\nu)} x^b q^n} \   \\ 
& = \left( \sum_{r \in \lambda + \mathbb{Z}} \, \sum_{s \in Z + \frac{1}{2} + \mathbb{Z}} + \sum_{r \in \lambda + \mathbb{Z}+\frac{1}{2}} \, \sum_{s \in Z + \mathbb{Z}} \right)\, 
x^r\, e^{2\pi i (\mu+\nu) s} \, \prod_{n=1}^{\infty} \frac{1}{\eta(\tau)^4}  \ ,
\end{aligned}
\ee
where $x=e^{2\pi i t}$ is associated to $J^{3}_0$ and we have used (\ref{lambda}). 
In the product we then restrict to the states with $Z_0=Z$, i.e.\ we only consider the terms that are proportional to $e^{2\pi i \mu Z}$. Since
\be\label{theta}
\begin{aligned}
\vartheta_3(\mu-\nu;2\tau) & = \sum_{n\in\mathbb{Z}} e^{2\pi i (\mu-\nu) n} q^{n^2} \ , \\ 
\vartheta_2(\mu-\nu;2\tau) & = \sum_{n\in\mathbb{Z}+\frac{1}{2}} e^{2\pi i (\mu-\nu) n} q^{n^2} \ ,
\end{aligned}
\ee%
see Appendix~\ref{app:theta} for our conventions, the relevant terms come from 
\begin{align}
& \sum_{r \in \lambda + \mathbb{Z}} \, \sum_{s \in Z + \frac{1}{2} + \mathbb{Z}} \, \sum_{n\in\mathbb{Z} + \frac{1}{2}} \, 
x^r\, e^{2\pi i (\mu+\nu) s} \,e^{2\pi i (\mu-\nu) n} q^{n^2} \,  \frac{\vartheta_3(z;2\tau) }{\eta(\tau)^6}  \\  
& + \sum_{r \in \lambda + \mathbb{Z}+\frac{1}{2}} \, \sum_{s \in Z + \mathbb{Z}}  \,  \sum_{n\in\mathbb{Z}} \, 
x^r\, e^{2\pi i (\mu+\nu) s} \,e^{2\pi i (\mu-\nu) n} q^{n^2} \,  \frac{\vartheta_2(z;2\tau) }{\eta(\tau)^6}\ .
 \end{align}
Picking up the term proportional to $e^{2\pi i \mu Z}$ then fixes $s$ to be equal to $s=Z-n$, and thus the character of ${\cal C}^{\lambda}_{Z,Z}$ equals 
 \begin{align}
 \text{ch}& \bigl[{\cal C}^{\lambda}_{Z,Z}\bigl]  (t,z,\nu,\mu;\tau) \\
 & = e^{2\pi i (\mu+\nu) Z}\,  \sum_{r \in \lambda + \mathbb{Z}} \, x^r\,  \frac{\vartheta_2(2\nu;2\tau) \vartheta_3(z;2\tau) }{\eta(\tau)^6}  \\ 
 & \qquad + 
e^{2\pi i (\mu+\nu) Z}\,  \sum_{r \in \lambda + \mathbb{Z}+\frac{1}{2}}\, x^r\,
\frac{\vartheta_3(2\nu;2\tau) \vartheta_2(z;2\tau)  }{\eta(\tau)^6} \ ,  \label{1.44}\\
& = e^{2\pi i (\mu+\nu) Z}\,   \sum_{r \in \lambda + \mathbb{Z} + \frac{1}{2}} \, x^r\,  \frac{\vartheta_2(2\nu+t;2\tau) \vartheta_3(z;2\tau) + \vartheta_3( 2\nu+t;2\tau) \vartheta_2(z;2\tau)}{\eta(\tau)^6}   \\
& = e^{2\pi i (\mu+\nu) Z}\,   \sum_{r \in \lambda + \mathbb{Z} + \frac{1}{2}} \, x^r\, \frac{\vartheta_2(\nu + \frac{t+z}{2};\tau) \vartheta_2(\nu + \frac{t-z}{2};\tau) }{\eta(\tau)^6} \ ,\label{1.46}
 \end{align}
where we have reassembled the sum over $n$ as in (\ref{theta}), and used that, because of the sum over $r$, we may add the chemical potential of $x=e^{2\pi i t}$ to the $\vartheta_3$, resp.\ the $\vartheta_2$, term. Finally, we have used (\ref{theta_2_theta_2}) in the last step. Note that relative to the calculation in \cite{Eberhardt:2018ouy} we have not removed the two $\eta$ factors associated to the $Z$ and $Y$ bosons. Indeed, the character is directly related to that of the representation $\mathscr{F}_{\lambda}$ of eq.~(\ref{F_lambda}) --- this can be calculated by the same methods as above, see also \cite{Eberhardt:2018ouy}
\begin{equation}\label{Frel}
\text{ch} \bigl[{\cal C}^{\lambda}_{Z,Z}\bigl]  (t,z,\nu,\mu;\tau) = \frac{e^{2\pi i (\mu+\nu) Z}\,}{\eta(\tau)^2} \, 
\text{ch} \bigl[\mathscr{F}_{\lambda} \bigl] (t,z,\nu;\tau)\ .
\end{equation}
Finally, the character of a general representation ${\cal C}^{\lambda}_{Y,Z}$ can be obtained from the above via (\ref{sigma_Z_flow}) 
\begin{equation}
\text{ch}\bigl[{\cal C}^{\lambda}_{Y,Z}\bigl]  (t,z,\nu,\mu;\tau) = e^{2\pi i \nu (Y-Z)} q^{Z(Z-Y)}  \text{ch}\bigl[{\cal C}^{\lambda}_{Z,Z}\bigl](t,z,\nu,\mu;\tau)   \ . 
\end{equation}
This just reflects that the value of $Y$ is in a sense a normal ordering choice, see the comment below eq.~(\ref{short_multiplet}). 
\medskip

For the following we will also need the character of the $\sigma$-spectrally flowed representations, where $\sigma$ was defined in (\ref{sigma}). This can be directly determined from the action on the Cartan generators as in (\ref{spec_flow_sigma}), and we thus find that 
\begin{align}
\text{ch}& \bigl[\sigma^w\bigl({\cal C}^{\lambda}_{Y,Z}\bigr)\bigr] (t,z,\nu,\mu;\tau) \\
& =  e^{2\pi i (\mu Z +\nu Y)}\,  q^{\frac{w^2}{2} + Z( Z- Y)}  e^{-2\pi i \nu w} \sum_{r \in \lambda + \mathbb{Z} + \frac{1}{2}} \, x^r\, q^{rw} \frac{\vartheta_2( \nu + \frac{t+z}{2} ;\tau) \vartheta_2( \nu + \frac{t-z}{2}  ;\tau) }{\eta(\tau)^6} \ ,
\end{align}
where we have used the theta function periodicity (\ref{theta_function_periodicity1}).

\subsection{Modular transformation}

For the study of the modular behaviour we pass as usual to the supercharacters, i.e.\ we introduce the operator $(-1)^{F}$ in the above characters and obtain
\be\label{u22_supercharacter}
\begin{aligned}    \text{sch}&\bigl[\sigma^{w}\bigl(\mathcal{C}^{\lambda}_{Y,Z}\bigl)\bigl](t,z,\nu,\mu;\tau) \\
    &=  e^{2\pi i (\mu Z +\nu Y)}\,  (-1)^w\, q^{\frac{w^2}{2} + Z(Z - Y)}\,  \sum_{r \in \lambda + \mathbb{Z} + \frac{1}{2}} \, e^{-2 \pi i \nu w}\, x^r\, q^{rw}\, \frac{\vartheta_1(  \frac{t+z}{2} + \nu ;\tau) \vartheta_1(\frac{t-z}{2} - \nu ;\tau) }{\eta(\tau)^6}  \\
    &= \frac{e^{2 \pi i(\mu Z + \nu Y)}}{\eta(\tau)^2} \, q^{Z(Z- Y)} \text{sch}\bigl[\sigma^{w}\bigl(\mathscr{F}_{\lambda }\bigl)\bigl](t,z,\nu;\tau)  \ ,
\end{aligned} 
\ee
where we have used (\ref{theta_2_theta_2}) and (\ref{theta_function_periodicity1}), as well as eq.~(\ref{Frel}) in the final step. In particular, its modular $S$-transformation therefore follows from the $S$-matrix for $\mathfrak{psu}(2|2)_1$, see \cite{Eberhardt:2018ouy}
\begin{equation}\label{S_transf_F}
\begin{aligned}    \text{sch}& \bigl[\sigma^{w}\bigl(\mathscr{F}_{\lambda}\bigl)\bigl]\bigl(\tfrac{t}{\tau},\tfrac{z}{\tau},\tfrac{\nu}{\tau};-\tfrac{1}{\tau}\bigl)  \\
& = e^{\frac{\pi i}{2 \tau}(-t^{2}+z^{2}+4\nu^{2})} \sum_{w' \in \mathbb{Z}} \int_{0}^{1} d\lambda' \, S^{\,\mathfrak{psu}}_{(w,\lambda),(w',\lambda')} \, \text{sch}\bigl[\sigma^{w'}\bigl(\mathscr{F}_{\lambda'}\bigl)\bigl](t,z,\nu;\tau) \ ,
\end{aligned}
\end{equation}
where we have used $\delta \left(\frac{a}{\tau} \right) = |a|$ for all $a\in\mathbb{R}$, and the modular $S$-matrix in (\ref{S_transf_F}) is given by
\begin{equation}\label{S_matrix}
    S^{\,\mathfrak{psu}}_{(w,\lambda),(w',\lambda')} = i \, \frac{|\tau|}{\tau} \, e^{2 \pi i \left(w'\lambda  + w\lambda' \right)} \ .
\end{equation}
Thus the modular $S$-matrix for $\mathfrak{u}(2|2)_1$ becomes 
\begin{equation}\label{S_tr_R_char}
\begin{aligned}   &\text{sch}\bigl[\sigma^{w}\bigl(\mathcal{C}^{\lambda}_{Y,Z}\bigl)\bigl]\bigl(\tfrac{t}{\tau},\tfrac{z}{\tau},\tfrac{\nu}{\tau},\tfrac{\mu}{\tau};-\tfrac{1}{\tau}\bigl)  \\
    & = e^{\frac{\pi i}{2 \tau}(-t^{2}+z^{2}-4\nu \mu)} \int_{\mathbb{R}^{2}} dY' \, dZ'  \sum_{w' \in \mathbb{Z}} \int_{0}^{1} d\lambda' \, S_{(Y,Z,w,\lambda),(Y',Z',w',\lambda')} \, \text{sch}\bigl[\sigma^{w'}\bigl(\mathcal{C}_{Z',Y'}^{\lambda'}\bigl)\bigl](t,z,\nu,\mu;\tau) \, ,
\end{aligned}
\end{equation}
where we recognise the prefactor $e^{\frac{\pi i}{2 \tau}(-t^{2}+z^{2}-4\nu \mu)}$ as being the usual factor associated to the chemical potentials. The modular $S$-matrix of $\mathfrak{u}(2|2)_1$ thus factorises into the $\widehat{\mathfrak{u}}(1)_{Y} \oplus \widehat{\mathfrak{u}}(1)_{Z}$ and $\mathfrak{psu}(2|2)_{1}$ modular $S$-matrices,
\begin{equation}\label{S_matrix_u22}
\begin{aligned}
S_{(Y,Z,w,\lambda),(Y',Z',w',\lambda')} &= i \, \frac{|\tau|}{\tau} \, e^{2 \pi i \left[-Z ( Y'-Z') -Z' ( Y-Z) + w'  \lambda  + w  \lambda'  \right]} \\
&= S^{\, \mathfrak{u}(1)\oplus\mathfrak{u}(1)}_{(Y,Z),(Y',Z')} \cdot S^{\,\mathfrak{psu}}_{(w,\lambda),(w',\lambda')}  
\ ,    
\end{aligned}
\end{equation}
where
\begin{equation}
S^{\, \mathfrak{u}(1)\oplus\mathfrak{u}(1)}_{(Y,Z),(Y',Z')} = e^{- 2 \pi i \left[-Z ( Y'-Z') -Z' ( Y-Z) \right]} \ . 
\end{equation}
The $S$-matrix for $\mathfrak{u}(2|2)_1$ is obviously symmetric, and it is (formally) unitary in the sense that
\begin{equation}\label{S_matr_unitarity}
\begin{aligned}
        \int_{\mathbb{R}^{2}} dY' \, dZ'  \sum_{w' \in \mathbb{Z}} \int_{0}^{1} d\lambda' \, & S^{\dagger}_{(Y,Z,w,\lambda),(Z',Y',w',\lambda')} \, S_{(Y',Z',w',\lambda'),(Z'',Y'',w'',\lambda'')}  \\
        & =  \delta(Y-Y'') \, \delta(Z-Z'') \, \delta_{w \, w''} \, \delta(\lambda-\lambda'' \,\text{mod}\,1) \,.
\end{aligned}
\end{equation}
Finally, the modular $T$-matrix of the supercharacters (\ref{u22_supercharacter}) is easily computed to be
 \begin{equation}  \text{sch}\bigl[\sigma^{w}\bigl(\mathcal{C}_{Y,Z}^{\lambda}\bigl)\bigl](t,z,\nu,\mu;\tau+1) = e^{Z(Z-Y)+\frac{w^{2}}{2}+(\lambda+\frac{1}{2})  w} \, \text{sch}\bigl[\sigma^{w}\bigl(\mathcal{C}_{Y,Z}^{\lambda}\bigl)\bigl](t,z,\nu,\mu;\tau) \ , 
 \end{equation}  
 which is also unitary in the sense of (\ref{S_matr_unitarity}).
In particular, the $T$-invariance therefore implies that the terms that appear in the modular invariant partition function must satisfy 
\begin{equation}\label{T_inv_cond}
     Z(Z-Y) + \lambda \, w = \bar{Z}( \bar{Z} -\bar{Y})  + \bar{\lambda} \, \bar{w} ~~\text{mod}\,1\ ,
\end{equation}
where the barred modes refer to the right-movers. 

\section{Modular invariants}\label{sec:modinv}
\setcounter{equation}{0}

Finally, we want to understand what modular invariants can be formed out of these characters. In particular, we are interested in modular invariants for which the $J^3_0$ eigenvalues are quantised since this is what should happen for the actual ${\rm U}(2|2)$ (and ${\rm PSU}(2|2)$) WZW model, see the comments in the paragraph above eq.~(\ref{Casimir}). 

Let us start though by writing down the `obvious' diagonal (resp.\ charge conjugation) modular invariant for $\mathfrak{u}(2|2)_1$, without worrying about the quantisation of the eigenvalues. This is simply obtained by taking the direct sum (or rather integral) over all of these representations, i.e.\
\begin{equation}\label{gl22_cc_spectrum}
    \mathcal{H}^{\mathfrak{u}(2|2)_{1}}_{\rm cc} = \OplusInt_{\mathbb{R}^{2}} dY \, dZ \,  \bigoplus_{w \in \mathbb{Z}} \OplusInt_{\mathbb{R}/\mathbb{Z}} d\lambda \, \sigma^{w}\bigl(\mathcal{C}^{\lambda}_{Y,Z}\bigl) \otimes \, \overline{\sigma^{-w}\bigl(\mathcal{C}^{-\lambda}_{-Y,-Z}\bigl)} \ ,
\end{equation}
and 
\begin{equation}\label{gl22_diagonal_spectrum}
    \mathcal{H}^{\mathfrak{u}(2|2)_{1}}_{\rm diag} = \OplusInt_{\mathbb{R}^{2}} dY \, dZ \, \bigoplus_{w \in \mathbb{Z}} \OplusInt_{\mathbb{R}/\mathbb{Z}} d\lambda \, \sigma^{w}\bigl(\mathcal{C}^{\lambda}_{Y,Z}\bigl) \otimes \, \overline{\sigma^{w}\bigl(\mathcal{C}^{\lambda}_{Y,Z}\bigl)}  \ . 
\end{equation}
The modular invariance of the charge conjugation modular invariant (\ref{gl22_cc_spectrum}) is simply a consequence of the unitarity of the modular $S$- and $T$-matrix, see eq.~(\ref{S_matr_unitarity}); the modular invariance of (\ref{gl22_diagonal_spectrum}) then follows from this since the $S$-matrix obeys the symmetry 
\begin{equation}\label{S_matr_symm}
    S_{(Y,Z,w,\lambda),(Y',Z',w',\lambda')} = S_{(-Y,-Z,-w,-\lambda),(-Y',-Z',-w',-\lambda')} \ .
\end{equation}
Actually, charge conjugation can be implemented independently on the pairs $(Y,\,Z)$ and $(w,\,\lambda)$, so that there are two additional modular invariant spectra. Furthermore, by removing the $(Y,Z)$ bosons and setting $Z_0=0$ we can also obtain the associated $\mathfrak{psu}(2|2)_1$ modular invariants from the above, and they take the form 
\begin{equation}
    \mathcal{H}^{\mathfrak{psu}(2|2)_{1}}_{\rm cc} = \bigoplus_{w \in \mathbb{Z}} \OplusInt_{\mathbb{R}/\mathbb{Z}} d \lambda \, \sigma^{w}\bigl(\mathscr{F}_{\lambda}\bigl) \otimes \, \overline{\sigma^{-w}\bigl(\mathscr{F}_{-\lambda}\bigl)} \ ,
\end{equation}
and
\begin{equation}\label{psu112_spectrum}
    \mathcal{H}^{\mathfrak{psu}(2|2)_{1}}_{\rm diag} = \bigoplus_{w \in \mathbb{Z}} \OplusInt_{\mathbb{R}/\mathbb{Z}} d \lambda \, \sigma^{w}\bigl(\mathscr{F}_{\lambda}\bigl) \otimes \, \overline{\sigma^{w}\bigl(\mathscr{F}_{\lambda}\bigl)} \ ,
\end{equation}
respectively, see \cite{Eberhardt:2018ouy}.

\subsection{Discrete spectra}

There are various ways in which discrete spectra may be obtained. For the application we have in mind, we are primarily interested in imposing the quantisation conditions\footnote{Note that the quantisation of $Z$ and $\lambda$ corresponds to quantising the $Z_0$- and $J^{3}_0$-eigenvalue respectively, while the free field realisation naturally yields $Z_0-Y_0 = 2V_0 \in \mathbb{Z}$, which corresponds to the other condition in (\ref{quantisation_conditions}).}
\begin{equation}
\label{quantisation_conditions}
      Z \in \tfrac{1}{2}\mathbb{Z} \ , ~~ Z=Y \ \text{mod}\,1 \ , ~~ \lambda = 0, \tfrac{1}{2} \ ,
 \end{equation}
 and analogously for the right-movers.
 We start by focusing on the representations with $\lambda = Z~\text{mod}\,1$. Because of $T$-invariance, see eq.~(\ref{T_inv_cond}), this then implies that 
\begin{equation}\label{T_inv}
     Z \, (w-Y) = \bar{Z} \, (\bar{w}-\bar{Y}) ~~\text{mod}\,1\ .
\end{equation}
Guided by (\ref{T_inv}), one finds the combinations
\begin{equation}\label{u22_invariants_R}
\begin{aligned}
    \bigoplus_{\substack{Z,\,\bar{Z} \in \mathbb{Z} }}  \, \bigoplus_{\substack{Y , \, \bar Y \in \mathbb{Z}}} \, \bigoplus_{w,\,\bar{w} \in \mathbb{Z} }    
&\sigma^{w}\bigl(\mathcal{C}_{Y,Z}^{0}\bigl) \otimes \, \overline{\sigma^{\bar{w}}\bigl(\mathcal{C}_{\bar{Y},\bar{Z}}^{0}\bigl)} \ ,  \\ 
 \bigoplus_{Z\in \frac{1}{2} \mathbb{Z}} \, \bigoplus_{w,\bar{w}\in \mathbb{Z}} \,  \bigoplus_{Z=\bar{Z}=Y=\bar Y \, \text{mod}\,1} \,     \bigoplus_{\substack{ Z+Y- \bar Z - \bar{Y} = \\ w-\bar{w}  \, \text{mod} \, 2 }} &   \sigma^{w}\bigl(\mathcal{C}_{Y,Z}^{Z}\bigl) \otimes \, \overline{\sigma^{\bar{w}}\bigl(\mathcal{C}_{\bar{Y},\bar{Z}}^{\bar{Z}}\bigl)} \ ,
  \end{aligned}
 \end{equation}
which are both modular invariant. Note that all of these representations are indecomposable, see eq.~(\ref{R_strucutre_diagr}). Thus the precise structure of the underlying vector space will actually involve their projective covers which are sketched in Appendix \ref{app_proj_cov}.

There is also another natural family of modular invariants for which $\lambda$ is not  necessarily equal to $Z$ mod $1$, and the simplest modular invariants (for which also the $Y$ and $Z$ eigenvalues are discrete) are 
 \begin{equation}\label{u22_invariants_Rp}
 \begin{aligned}
      \bigoplus_{Z, \, \bar{Z} \in \mathbb{Z} }\, \bigoplus_{Y,\,\bar{Y} \in \mathbb{Z} }  \,
 \bigoplus_{w=\bar{w} \, \text{mod}\,2} \, 
 \bigoplus_{\lambda = 0, \frac{1}{2} } \, & \sigma^{w}\bigl(\mathcal{C}_{Y,Z}^{\lambda}\bigl) \otimes \, \overline{\sigma^{\bar{w}}\bigl(\mathcal{C}_{\bar{Y},  \bar{Z}}^{\lambda} \bigl)}  \ , \\
\bigoplus_{Z = \bar{Z} \text{mod\,1}} \, \bigoplus_{\substack{Y = \bar Y \, \text{mod}\,1, \\ Z+Y =\bar Z + \bar{Y}  \, \text{mod}\,2}}\, 
\bigoplus_{w=\bar{w} \, \text{mod}\,2} \, 
 \bigoplus_{\lambda = 0, \frac{1}{2}} \, & \sigma^{w}\bigl(\mathcal{C}_{Y,Z}^{\lambda}) \otimes \, \overline{\sigma^{\bar{w}}\bigl(\mathcal{C}_{\bar{Y}, \, \bar{Z}}^{\lambda} \bigl)}  \ . 
 \end{aligned}
 \end{equation}
Given that the $S$-matrix of $\mathfrak{u}(2|2)_1$ factorises into the $S$-matrix of $\mathfrak{psu}(2|2)_1$, as well as the $S$-matrix of the $(Y,Z)$ boson system, see eq.~(\ref{S_matrix_u22}), we can also associate $\mathfrak{psu}(2|2)_1$ modular invariants to these expressions, and the first modular invariant in eq.~(\ref{u22_invariants_R}) leads to 
\begin{equation}\label{psu22_spectrum}
    \mathcal{H}^{\mathfrak{psu}(2|2)_{1}}_{1} = \bigoplus_{w,\, \bar{w} \in \mathbb{Z}} \sigma^{w}\bigl(\mathscr{F}_{0}\bigl) \otimes \, \overline{\sigma^{\bar{w}}\bigl(\mathscr{F}_{0}\bigl)} \ ,  
\end{equation}
while the other two in eq.~(\ref{u22_invariants_R}) and those in eq.~(\ref{u22_invariants_Rp}) give rise to 
\begin{equation}\label{psu22_spectrum_2}
    \mathcal{H}^{\mathfrak{psu}(2|2)_{1}}_{2} = \bigoplus_{w=\bar{w} \, \text{mod} \, 2} \, \bigoplus_{\lambda = 0, \, \frac{1}{2}} \sigma^{w}\bigl(\mathscr{F}_{\lambda}\bigl) \otimes \, \overline{\sigma^{\bar{w}}\bigl(\mathscr{F}_{\lambda}\bigl)}  \ .
\end{equation}
Note that we can think of (\ref{psu22_spectrum}) as arising from a $\mathbb{Z}$ orbifold of (\ref{psu112_spectrum})
\begin{equation}
    \mathcal{H}^{\mathfrak{psu}(2|2)_{1}}_{1} \cong \mathcal{H}^{\mathfrak{psu}(2|2)_{1}}_{\rm diag} \bigl/ \mathbb{Z} \ ,
\end{equation}
where the orbifold action is defined by 
\begin{equation}
    m  \mapsto e^{2 \pi i m J^{3}_{0}} ~~~~ \text{for}~ m \in \mathbb{Z} \ .
\end{equation}
In particular, in the untwisted sector we therefore project onto states for which the $J^3_0$-eigenvalue is an integer, while the $m$'th twisted sector consists of the states for which $w-\bar{w} = m$. It is also worth mentioning that (\ref{psu22_spectrum}) has in fact the simple product form 
\begin{equation}\label{5.14}
 \mathcal{H}^{\mathfrak{psu}(2|2)_{1}}_{1} = \mathbb{F} \otimes \overline{\mathbb{F}} \ , \qquad \hbox{with} \qquad \mathbb{F} =  \bigoplus_{w\in \mathbb{Z}} \sigma^{w}\bigl(\mathscr{F}_{0}\bigl) \ . 
 \end{equation}

\subsection{Additional Discrete modular invariants}

As we saw in Section~\ref{sect_free_field_repr}, the representations  $\mathcal{C}^{Z}_{Y,Z}$ for $Z \in \tfrac{1}{2}\mathbb{Z}$ are not irreducible, but have an interesting composition series which we worked out in eq.~(\ref{R_strucutre_diagr}). In terms of (super-)characters this composition series implies that the characters can be expressed in terms of sums of characters,\footnote{The more formal statement is that this is true on the level of the Grothendieck ring, see \cite[(4.12)]{Eberhardt:2018ouy} for more details.}  i.e.\ that 
\begin{equation}
\begin{aligned}
\text{sch} \bigl[  \mathcal{C}^{Z}_{Y,Z} \bigr] &=  \, \text{sch} \bigl[ \rho^{2Z} \sigma_Z^{Z-Y+1} \bigl( \mathcal{H}_{0,0} \bigl)  \bigl] + \, 
\text{sch} \bigl[  \sigma \rho^{2Z} \sigma_Z^{Z-Y} \bigl( \mathcal{H}_{0,0} \bigl)  \bigl] \\ 
 & \qquad +  \text{sch} \bigl[ \sigma^{-1} \rho^{2Z} \sigma_Z^{Z-Y} \bigl( \mathcal{H}_{0,0} \bigl) \bigl] + \,
\text{sch} \bigl[ \rho^{2Z} \sigma_Z^{Z-Y-1} \bigl( \mathcal{H}_{0,0} \bigl)  \bigl]   \ .
\end{aligned} 
\end{equation}
It follows that the two partition functions of (\ref{u22_invariants_R}) are proportional to the combinations 
\begin{equation}\label{u22_invariants_H}
\begin{aligned}
 \bigoplus_{r,\,\bar{r} , \, s , \, \bar s , \, w,\,\bar{w} \, \in \mathbb{Z} } & \sigma^{w} \rho^{2r}\sigma_{Z}^{s} \bigl(\mathcal{H}_{0,0}\bigl) \otimes \,\overline{\sigma^{\bar{w}}\rho^{2\bar r}\sigma_{Z}^{\bar s}\bigl(\mathcal{H}_{0,0}\bigl)} \ , \\ 
 \bigoplus_{\substack{r=\bar{r} \, \text{mod}\,2 \\ w+s=\bar w + \bar s \, \text{mod}\,2}} & \sigma^{w} \rho^{r}\sigma_{Z}^{s} \bigl(\mathcal{H}_{0,0}\bigl) \otimes \,\overline{\sigma^{\bar{w}}\rho^{\bar r}\sigma_{Z}^{\bar s}\bigl(\mathcal{H}_{0,0}\bigl)} \ ,\\
 \end{aligned}
 \end{equation}
 respectively. 
 Moreover, we can similarly apply the same argument to eq.~(\ref{psu22_spectrum}) using the composition series of \cite[(4.14)]{Eberhardt:2018ouy}, and thereby conclude that the combination 
 \begin{equation}\label{psu22_inv}
     \bigoplus_{w,\,\bar{w} \, \in \mathbb{Z}} \sigma^{w}\bigl(\mathscr{L}\bigl) \otimes \, \overline{\sigma^{\bar w}\bigl(\mathscr{L}\bigl)} \ 
 \end{equation}
is also modular invariant. Note that (\ref{psu22_inv}) can also be obtained from the first invariant in (\ref{u22_invariants_H}) by quotienting out the $(Y,Z)$ boson system. 

While these considerations do not prove that these combinations define consistent CFTs, it is maybe worth pointing out that eqs.~(\ref{u22_invariants_H}) and (\ref{psu22_inv}) have a nice interpretation in terms of extended algebras. For example, we can write (\ref{psu22_inv}) at least formally as 
\begin{equation}\label{L_invariant}
    \mathbb{L} \otimes \overline{\mathbb{L}} \ , \qquad \hbox{where} \quad 
    \mathbb{L} = \bigoplus_{w\in \mathbb{Z}} \sigma^{w}\bigl(\mathscr{L}\bigl) \ .
\end{equation}
Furthermore, since $\mathscr{L}$ is the identity representation of $\mathfrak{psu}(1,1|2)_1$, it has trivial fusion rules, and one would expect that the fusion of its spectrally flowed images also close among themselves. This is also supported by the observation that all the states in $\sigma^{w}\bigl(\mathscr{L}\bigl)$ have in fact integer conformal dimension. However, some of these conformal dimensions are actually negative, and thus $\mathbb{L}$ must be a somewhat unusual chiral algebra (if it exists).\footnote{Chiral algebras that contain fields of negative conformal dimension have recently also appeared in discussions of celestial holography, see e.g.\ \cite{Raclariu:2021zjz}.} Note that the same discussion applies also to (\ref{5.14}), except that $\mathcal{L}$ is replaced by $\mathscr{F}_0$. 
Similarly, we can rewrite the first invariant in (\ref{u22_invariants_H}) as 
\begin{equation}\label{rewrite_H_inv}
\begin{aligned}
       &\mathbb{H} \otimes \overline{\mathbb{H}} \ , \qquad \qquad \qquad \qquad \text{where} \quad \mathbb{H} = \bigoplus_{Y,Z,w \in \mathbb{Z}} \sigma^{w} \bigl(\mathcal{H}_{Y,Z} \bigl) \ , 
\end{aligned}
\end{equation}
which is the analogue of (\ref{L_invariant}) --- indeed, this modular invariant can be obtained from the latter by quotienting out by $(Y,Z)$. 

Furthermore, the second invariant in (\ref{u22_invariants_H}) can also be thought of as a free field invariant. 
To this end we consider the Ramond sector representation $\mathrm{R}$ defined by (\ref{sympl_bos_zero}) and (\ref{ferm_zero_modes}) with $m_1,m_2 \in \tfrac{1}{2} \mathbb{N}_0$, which decomposes into $\mathfrak{u}(2|2)_1$ representations as 
\begin{equation}\label{R+_decomposition}
    \mathrm{R} \cong \bigoplus_{Z \in \frac{1}{2}\mathbb{Z}} \mathcal{H}_{Z+1,Z}  \cong \bigoplus_{r \in \mathbb{Z}} \rho^{r}\sigma_{Z}^{-1} \bigl(\mathcal{H}_{0,0} \bigl) \ . 
\end{equation}
Thus the invariant in the second line of (\ref{u22_invariants_H}) is equivalent to the free field invariant
\begin{equation}\label{free_field_inv}
    \bigoplus_{\substack{w,s,\bar{w},\bar{s}\in \mathbb{Z}:\\ w = \bar{w} \, \text{mod}\,2, \\ s = \bar{s} \, \text{mod}\,2}} \sigma^{w} \sigma_Z^{s} \bigl(\mathrm{R} \bigl) \otimes \, \overline{\sigma^{\bar w} \sigma_Z^{\bar s} \bigl(\mathrm{R} \bigl)} \ .
\end{equation}

\section{Conclusions}\label{sec:concl}
\setcounter{equation}{0}

In this paper we have studied the WZW model based on the super Lie algebras $\mathfrak{u}(2|2)_1$ and 
$\mathfrak{psu}(2|2)_1$. In particular, we have classified the most general highest weight representations, and we have found free field realisations for them. This has allowed us to find a number of modular invariant partition functions. Quite remarkably, some of them seem to be maximally extended modular invariants, see eqs.~(\ref{L_invariant}) and (\ref{rewrite_H_inv}). 

One of the main motivations for this work was to find other candidate worldsheet theories that could describe the string background that is dual to free ${\cal N}=4$ SYM in 4D, see \cite{Gaberdiel:2021qbb,Gaberdiel:2021jrv}. While none of the above theories seem to reproduce directly the correct spectrum, we suspect that they will nevertheless play an important role in that context. 

The $\mathfrak{psu}(1,1|2)_1$ WZW model that describes the string background dual to the symmetric orbifold of $\mathbb{T}^4$ \cite{Eberhardt:2018ouy,Eberhardt:2019ywk} has quite remarkable properties --- in particular, its correlators localise to configurations that admit holomorphic covering maps \cite{Eberhardt:2019ywk,Dei:2020zui}. It would be interesting to study whether a similar phenomenon also holds for the above $\mathfrak{u}(2|2)_1$ theories. In particular, given that the left- and right-moving spectral flows are uncorrelated for the modular invariants we have found --- we have independent sums over $w$ and $\bar{w}$ ---  the nature of the localisation differs from what happened for AdS$_3$. In fact, this may be required to account for the correct structure of the correlators of ${\cal N}=4$ SYM in 4D. 

\section*{Acknowledgements}

This paper is based on the Master thesis of one of us (EM). We thank Rajesh Gopakumar and David Ridout for useful conversations. MRG is supported in part by the Simons Foundation grant 994306  (Simons Collaboration on Confinement and QCD Strings). The work of the group is furthermore supported by a personal grant from the Swiss National Science Foundation, as well as the NCCR SwissMAP that is also funded by the Swiss National Science Foundation.

\appendix

\section{Shortening of $\mathfrak{u}(2|2)$-multiplets}
\setcounter{equation}{0}
\label{appendix_shortening}
\renewcommand{\theequation}{A.\arabic{equation}}

In this appendix we prove the shortening condition for the multiplet (\ref{full_n_1}) for the case of $j$ labelling continuous $\mathfrak{su}(2)$ representations; the other cases can be deduced from that by considering subrepresentations (or quotient representations), or by an analogous analysis. In the following we will omit the zero mode labels, since we are only considering the finite Lie superalgebra $\mathfrak{u}(2|2)$. 

Let us work with the convention that the state $|m\,,0 \rangle \in ( C_j^{\lambda} \,, \mathbf{1} )$ at the top of the diagram in eq.~(\ref{full_n_1}) is annihilated by the fermionic generators 
\begin{equation}
    S^{\alpha \beta +} \, |m\,,0 \rangle = 0 ~~~~~~\text{for}~\alpha, \beta \in\{+,-\} \  .
\end{equation}
We need to impose the condition that the summands $( C_j^{\lambda} \,, \mathbf{3} )$ in the middle line of (\ref{full_n_1}) drop out. They are obtained from $|m\,,0 \rangle$ upon applying 
\begin{equation}
\mathcal{N}(m) = S^{-+-} S^{++-} \, |m\,,0 \rangle \ , 
\end{equation}
i.e.\ $\mathcal{N}(m)$ must be a null-vector. This amounts to requiring that 
\begin{equation}\label{n_1_z_condition}
    S^{--+}S^{+-+} \, \mathcal{N}(m) = \bigl( Z(Z+1) - j(j+1) \bigl) \, |m\,,0\rangle \ ,
\end{equation}
which fixes the value of $j$ to either $Z$ or $-Z-1$. Then, the vanishing of (\ref{n_1_z_condition}) implies that $\mathcal{N}(m)$ is null, and hence that all the states obtained from it by the application of fermionic raising operators are also null. The only non-trivial such states are
\begin{equation}\label{n_1_null_vectors}
\begin{aligned}
    S^{+-+} \, \mathcal{N}(m) &= -\bigl((Z+m+1) \, S^{++-} + S^{-+-}J^{+} \bigl)\, |m\,,0\rangle \ ,\\
    S^{--+}\, \mathcal{N}(m) &= \bigl((Z-m+1) \, S^{-+-}  + S^{++-} J^{-} \bigl) \, |m\,,0\rangle \ .
\end{aligned}
\end{equation}
We set $Z = j$, which by the symmetry in $j$ is identical to $j=-Z-1$. The states in (\ref{n_1_null_vectors}) being null then translates into
    \begin{equation}\label{n_1_null_vectors1}
        S^{++-} \, |m\,,0\rangle \,, ~ S^{-+-} \, |m\,,0\rangle \in \bigl(C^{\lambda + \frac{1}{2}}_{j-\frac{1}{2}} , \mathbf{2} \bigl) \ .
    \end{equation}
   It follows that
    \begin{equation}\label{S___}
        S^{\mp --} \, |m\,,0\rangle = [K^{-}, S^{\mp +-}] \, |m\,,0\rangle = K^{-}S^{\mp +-} \, |m\,,0\rangle  \in \bigl(C^{\lambda + \frac{1}{2}}_{j-\frac{1}{2}} , \mathbf{2} \bigl) \ ,
    \end{equation}
    and so there is only one additional representation in the Clifford module, which is $( C_{j - 1}^{\lambda} \,, \mathbf{1})$, and which is generated by $S^{---} S^{-+-} \ |m\,0\rangle$. We have thus proven the shortening of (\ref{full_n_1}), which again gives the multiplet (\ref{short_multiplet}).
    
For what concerns (\ref{full_n_2}), it is easy to check that there is no solution such that both representations $\mathbf{3}$ in the second line drop out, and thus there are no additional highest weight representations of $\mathfrak{u}(2|2)_1$.

\subsection{Other short multiplets}

For completeness let us also describe explicitly the short $\mathfrak{u}(2|2)$ multiplets involving the discrete and finite dimensional representations of $\mathfrak{su}(2)$. For the discrete case we have 
    \begin{equation}\label{short_n_1_highest_discrete}
\begin{aligned}  
    &\big( D^{\pm}_{Z} \, , \mathbf{1} \big)_{Y \pm 1,\,Z}    \oplus
    \big( D^{\pm}_{Z \mp \frac{1}{2}} \, , \mathbf{2} \big)_{Y,\,Z} \oplus \big( D^{\pm}_{Z \mp 1} \, , \mathbf{1} \big)_{Y\mp 1,\,Z}\,~~~~~~~~~~~~~~~\,~~~~~~~~\text{for} ~Z  \notin \tfrac{1}{2}\mathbb{Z}  \ , \\
    &\big( D^{\pm}_{ \mp Z} \, , \mathbf{1} \big)_{Y-1,\,Z}   \oplus
    \big( D^{\pm}_{ \mp(Z + \frac{1}{2})} \, , \mathbf{2} \big)_{Y,\,Z} \oplus \big( D^{\pm}_{ \mp(Z + 1)} \, , \mathbf{1} \big)_{Y +\, 1 ,\,Z} \, \qquad \qquad \quad \text{for} ~Z  \in \tfrac{1}{2}\mathbb{Z}_{>0}  \ , \\ 
        &\big( D^{\pm}_{ \pm Z} \, , \mathbf{1} \big)_{Y+1,\,Z}   \oplus
    \big( D^{\pm}_{ \pm(Z - \frac{1}{2})} \, , \mathbf{2} \big)_{Y,\,Z} \oplus \big( D^{\pm}_{ \pm(Z - 1)} \, , \mathbf{1} \big)_{Y -1  ,\,Z} \, \qquad \qquad \quad \text{for} ~Z  \in \tfrac{1}{2}\mathbb{Z}_{<0}  \ , \\ 
     &\big( D^{\pm}_{\mp1} \, , \mathbf{1} \big)_{Y+1,\,0}  \oplus
     \big(  D^{\pm}_{\mp \frac{1}{2}} \, , \mathbf{2} \big)_{Y,\,0} \oplus
    \big(  D^{\pm}_{\mp1} \, , \mathbf{1} \big)_{Y-1,\,0} ~~~~~~~~~~~~~~~~~~~~~~~\,~~~~\,\text{for}~Z=0 \ ,
\end{aligned}
\end{equation}
and we denote the affine $\mathfrak{u}(2|2)_1$ representations corresponding to these multiplets by $\mathcal{D}^{\pm}_{Y,Z}$. Then, 
\begin{equation}\label{conj_on_D}
    \mathcal{D}^{\pm}_{Y,Z} = \bigl( \mathcal{D}^{\mp}_{-Y,-Z} \bigl)^{*} ~~~~ \forall \, Y,Z \in \mathbb{R} \ .
\end{equation} 
The short multiplets involving the finite-dimensional representations $\mathcal{H}_{Y,Z}$ are 
\begin{equation}\label{short_n_1}
\begin{aligned} 
  &\big( H_{Z} \, , \mathbf{1} \big)_{Y ,\,Z}   \oplus
   \big(H_{Z - \frac{1}{2}} \, , \mathbf{2} \big)_{Y -1 ,\,Z} \oplus \big(H_{Z - 1} \ , \mathbf{1} \big)_{Y - 2,\,Z}~\, \qquad \quad\ \text{for} ~ Z \in \tfrac{1}{2}\mathbb{Z} \ , Z \geq 1 \ , \\
  &\big( H_{-Z} \, , \mathbf{1} \big)_{Y ,\,Z}   \oplus
   \big(H_{-Z - \frac{1}{2}} \, , \mathbf{2} \big)_{Y +1 ,\,Z} \oplus \big(H_{-Z - 1} \ , \mathbf{1} \big)_{Y + 2,\,Z}~\, \quad  \ \   \text{for} ~ Z \in \tfrac{1}{2}\mathbb{Z} \ , Z \leq -1 \ , \\
  &\big( H_{\frac{1}{2}} \, , \mathbf{1} \big)_{Y  ,\,Z} \oplus \big( H_0 \, , \mathbf{2} \big)_{Y  \mp 1,\,Z}~~~~~~~~~~~~~~~~~~~~~~~~~~~\,~~~~~~~~~~ \ \  \,\,\text{for}\, Z= \pm \tfrac{1}{2}   \ , \\
  &\big( H_0 \, , \mathbf{1} \big)_{Y,\,0}~~~~~~~~~~~~~~~~~~~~~~~~~~~~~~~~~~~~~~~~~~~~~~~~~~~~~~~~~~~~~~ \,\text{for}~Z=0\ ,
\end{aligned}
\end{equation}
and we denote the corresponding affine representations by $\mathcal{H}_{Y,Z}$. Then, 
\begin{equation}\label{conj_on_H}
    \mathcal{H}_{-Y,-Z} = \bigl(\mathcal{H}_{Y,Z}\bigl)^{*} ~~~~ \forall \, Y \in \mathbb{R} \,,  Z \in \tfrac{1}{2}\mathbb{Z} \ .
\end{equation}

\section{Projective covers}\label{app_proj_cov}
\setcounter{equation}{0}
\renewcommand{\theequation}{B.\arabic{equation}}

For the actual WZW model spectrum on Lie supergroups the determination of the projective covers\footnote{It is known that the projective cover of a typical module is the module itself, so this notion is only important for the atypical modules, see \cite{Quella:2007hr}. However, for $k=1$ all modules are atypical, and thus this is the `generic' case.} is of importance \cite{Quella:2007hr}, see also \cite{Gaberdiel:2007jv}, since the atypical part of the spectrum consists of projective covers quotiented by an ideal that makes the action of $L_{0} - \bar{L}_{0}$ diagonalisable \cite{Gaberdiel:1998ps}.

The vacuum module of an affine Lie superalgebra is always atypical \cite{Quella:2007hr}. This is the case for instance, for the $\mathfrak{psu}(2|2)_1$ vacuum $\mathscr{L}$, which is part of an indecomposable representation $\mathscr{F}_{0}$, see \cite[(4.14)]{Eberhardt:2018ouy}. Both of these representations have vanishing Casimir and are covered by the same projective module $\mathscr{T}$, whose structure is given in \cite[(C.27)]{Eberhardt:2018ouy}.

For what concerns the $\mathfrak{u}(2|2)_{1}$ indecomposables, for $Z \in \tfrac{1}{2}\mathbb{Z}$ and $Y \in \mathbb{R}$ we denote by $\mathscr{T}_{Y,Z}$ the projective cover of $\mathcal{H}_{Y,Z}$, which by (\ref{R_strucutre_diagr}) is the same as that of $\mathcal{C}^{Z}_{Y \pm 1,Z}$. By functoriality and (\ref{spec_flow_iso_rho}) we have that
\begin{equation}\label{T_YZ}
    \mathscr{T}_{Y,Z} \cong \sigma_Z^{Z-Y} \bigl(\mathscr{T}_{Z,Z} \bigl) \cong \sigma_Z^{Z-Y} \circ \rho^{2Z}\bigl(\mathscr{T}_{0,0}\bigl) \ ,
\end{equation}
and thus the structure of $\mathscr{T}_{Y,Z}$ can be obtained from that of $\mathscr{T}_{0,0}$ in \cite{Eberhardt:2018ouy},
\begin{equation}\label{TYZ_struct}
   \begin{tikzcd}[sep=small]
        & &  \mathcal{H}_{Y,Z} \arrow[dl] \arrow[dr] \\
        & \sigma\bigl(\mathcal{H}_{Y \pm 1,Z}\bigl)  \arrow[dl]  \arrow[dr] & &  \sigma^{-1}\bigl(\mathcal{H}_{Y \pm 1,Z}\bigl)  \arrow[dl]  \arrow[dr] \\
        \sigma^{2}\bigl(\mathcal{H}_{Y,Z}\bigl)  \arrow[dr] & &  2 \, \mathcal{H}_{Y,Z} \oplus \mathcal{H}_{ Y \pm 2,Z} \arrow[dl] \arrow[dr] & & \sigma^{-2}\bigl(\mathcal{H}_{Y, Z}\bigl) \arrow[dl]  \\
        & \sigma\bigl(\mathcal{H}_{Y \pm 1,Z}\bigl)    \arrow[dr] & &  \sigma^{-1}\bigl(\mathcal{H}_{Y \pm 1,Z}\bigl)  \arrow[dl]   \\
        & &  \mathcal{H}_{Y,Z}  \ ,
\end{tikzcd} 
\end{equation}
where we have spelled out the case for $Z \in \tfrac{1}{2}\mathbb{Z}$, and all the terms (with the different signs) are present, e.g.\ in the first line there are four representations. 
In particular, the analogue of \cite[(C.28)]{Eberhardt:2018ouy} is 
\begin{equation}
    \mathscr{T}_{Y,Z} \sim \sigma^{-1}\bigl(\mathcal{C}^{Z}_{Y,Z}\bigl) \, \oplus \, \mathcal{C}^{Z}_{Y-1,Z} \oplus \, \mathcal{C}_{Y+1,Z}^{Z} \oplus \, \sigma \bigl(\mathcal{C}_{Y,Z}^{Z} \bigl) ~~~~ \forall \, Y \in \mathbb{R} \, , Z \in \tfrac{1}{2}\mathbb{Z} \ ,
\end{equation}
which can be read off from (\ref{TYZ_struct}).

\section{Theta function identities}\label{app:theta}
\setcounter{equation}{0}
\renewcommand{\theequation}{C.\arabic{equation}}

Here we summarise the theta function identities we need in the main part of the text. We define the theta functions as

\begin{equation}
\begin{aligned}
       \vartheta \begin{bmatrix} \alpha \\ \beta
               \end{bmatrix} (z;\tau) & := \sum_{n\in \mathbb{Z}} e^{\pi i (n+\alpha)^{2}\tau + 2\pi i (n+\alpha )(z + \beta)} \\
               &= e^{2 \pi i  \alpha(z + \beta) } q^{\frac{\alpha^{2}}{2}} \prod_{n = 1}^{\infty}(1-q^{n})(1+q^{n + \alpha - \frac{1}{2}} e^{2 \pi i (z + \beta )})(1 + q^{n- \alpha -\frac{1}{2}} e^{-2 \pi i (z + \beta)}) \ ,
\end{aligned}
\end{equation}
where the second equality holds by applying the \textit{Jacobi triple product}. The Jacobi theta functions are then
\begin{equation}
    \vartheta_{1} := \vartheta \begin{bmatrix} \frac{1}{2} \\ \frac{1}{2}
               \end{bmatrix} \ ,~~~~ \vartheta_{2} := \vartheta \begin{bmatrix} \frac{1}{2} \\ 0
               \end{bmatrix} \ ,~~~~ \vartheta_{3} := \vartheta \begin{bmatrix} 0 \\ 0
               \end{bmatrix} \ ,~~~~ \vartheta_{4} := \vartheta \begin{bmatrix} 0 \\ \frac{1}{2}
               \end{bmatrix} \ .
\end{equation}
These functions obey the addition rules
\begin{equation}\label{theta_2_theta_2}
\begin{aligned}
        \vartheta_{1}(\tfrac{z+t}{2};\tau)\vartheta_{1}(\tfrac{z-t}{2};\tau) &= \vartheta_{2}(z;2\tau) \vartheta_{3}(t;2\tau)- \vartheta_{3}(z;2\tau) \vartheta_{2}(t;2\tau) \ , \\  
        \vartheta_{2}(\tfrac{z+t}{2};\tau)\vartheta_{2}(\tfrac{z-t}{2};\tau) &= \vartheta_{2}(z;2\tau) \vartheta_{3}(t;2\tau) + \vartheta_{3}(z;2\tau) \vartheta_{2}(t;2\tau) \ , \\        
\end{aligned}
\end{equation}
and the quasi-periodicity relations
\begin{equation}
\label{theta_function_periodicity1}
\begin{aligned}
     \vartheta_{1}(z + w \tau;\tau) &= (-1)^{w} e^{- 2 \pi i w z}q^{\frac{-w^2}{2}} \vartheta_{1}(z;\tau) \ ,\\
    \vartheta_{2}(z + w \tau;\tau) &= e^{- 2 \pi i w z}q^{\frac{-w^2}{2}} \vartheta_{2}(z;\tau)\ , \\
\end{aligned}
\end{equation}
for every $w \in \mathbb{Z}$.
We also need the modular transformations
\begin{equation}
\begin{aligned}
\vartheta_{1}(z;\tau+1) &= e^{\frac{\pi i}{4}} \vartheta_{1}(z;\tau) \ , \\
\vartheta_{1}(\tfrac{z}{\tau}; -\tfrac{1}{\tau}) &= - i \sqrt{-i \tau}\,  e^{\frac{\pi i z^{2}}{\tau}} \vartheta_{1}(z;\tau) \ . 
\end{aligned}
\end{equation}
Finally, we recall the Dedekind eta function
\begin{equation}
    \eta(\tau) = q^{\frac{1}{24}} \prod_{n=1}^{\infty}(1-q^{n}) \ ,
\end{equation}
and its modular transformations 
\begin{equation}\label{eta_modular_tr}
    \begin{aligned}
        \eta(\tau + 1) &= e^{\frac{\pi i }{12}} \, \eta(\tau) \ , \\
        \eta(-\tfrac{1}{\tau}) &= \sqrt{-i \tau} \, \eta(\tau) \ . \\
    \end{aligned}
\end{equation}

\end{document}